\documentclass[fleqn,usenatbib]{mnras}

\usepackage{newtxtext,newtxmath}
\usepackage{float}
\usepackage{booktabs}

\usepackage[T1]{fontenc}

\DeclareRobustCommand{\VAN}[3]{#2}
\let\VANthebibliography\thebibliography
\def\thebibliography{\DeclareRobustCommand{\VAN}[3]{##3}\VANthebibliography}

\usepackage{graphicx}	
\usepackage{amsmath}	
\usepackage{comment}
\usepackage{bm}
\usepackage{longtable}
\usepackage{multirow}
\usepackage{subfig}
\usepackage{fixltx2e}

\title[Origin of X-rays in HAeBe stars]{Chandra X-ray Analysis of Herbig Ae/Be Stars}

\author[Hema Anilkumar et al.]{
Hema Anilkumar,$^{1}$\thanks{E-mail: hema.anilkumar@res.christuniversity.in}
Blesson Mathew,$^{1}$
V. Jithesh,$^{1}$
Sreeja S. Kartha,$^{1}$
P. Manoj,$^{2}$
Mayank Narang$^{3}$
\newauthor and Mahathi Chavali,$^{4}$
\\
$^{1}$Department of Physics \& Electronics, CHRIST (Deemed to be University), Bangalore 560029, India\\
$^{2}$Department of Astronomy and Astrophysics, Tata Institute of Fundamental Research Homi Bhabha Road, Colaba, Mumbai 400005, India\\
$^{3}$Academia Sinica Institute of Astronomy \& Astrophysics, 11F of Astro-Math Bldg., No. 1, Sec. 4, Roosevelt Road, Taipei 10617, Taiwan, Republic of China\\
$^{4}$Max Planck Institute for Radio Astronomy, Auf dem Hügel 69, 53121 Bonn, Germany 
}

\date{Accepted XXX. Received YYY; in original form ZZZ}

\pubyear{2023}

\begin{document}
\label{firstpage}
\pagerange{\pageref{firstpage}--\pageref{lastpage}}
\maketitle

\begin{abstract}
Herbig Ae/Be (HAeBe) stars are intermediate-mass pre-main sequence stars, characterized by infrared excess and emission lines. They are observed to emit X-rays, whose origin is a matter of discussion and not settled yet. X-ray emission is not expected in HAeBe stars, as they lack the sub-surface convective zone. In this study, we retrieved observations from the \textit{Chandra} archive for 62 HAeBe stars, among which 44 sources (detection fraction $\sim$71\%) were detected in X-rays, with 7 being new detections. We use this sample as a test bed to conduct a comparative analysis of the X-ray properties of HAeBe stars and their low-mass counterparts, T Tauri Stars (TTSs). Further, we compare the X-ray properties of HAeBe stars and TTSs with optical and IR properties to constrain the X-ray emission mechanism in HAeBe stars. We found no correlation between X-ray emission and disk properties of HAeBe stars, confirming that X-rays are not related to accretion shocks. About 56\% of HAeBe stars without any known sub-arcsec companions have lower plasma temperatures (kT $\leq$ 2\,keV). We observe flaring/variability in HAeBe stars with confirmed low-mass companions. These stars show plasma temperatures > 2\,keV, similar to TTSs. Guided by this information we discuss the role of a T Tauri companion for X-ray emission seen in our sample of HAeBe stars. From the results obtained in this paper, we suggest that X-ray emission from HAeBe stars may not be related to accretion shocks or hidden TTS, but rather can be due to magnetically driven coronal emission.
\end{abstract}




\section{Introduction}\label{section_1:intro}
Herbig Ae/Be (HAeBe) stars are intermediate mass (2\,M$_\mathrm{\odot}$ $\leq$ M$_\mathrm{\star}$ $\leq$ 8\,M$_\mathrm{\odot}$) young stellar objects. They are the higher mass counterparts of the T Tauri stars (TTS) \citep[e.g.,][]{2023Brittain}. Like TTSs, HAeBe stars also show H$\alpha$ emission and infrared (IR) excess from a dusty circumstellar material. In addition to this, they are also observed to emit UV and X-rays \citep{ZnP1994, Damiani1994, Malfait1998}. The high energy radiation in low-mass main-sequence (MS) stars and pre-main-sequence (PMS) stars is produced predominantly due to the involvement of the magnetic field. However, the subsurface convection zone required for the production of magnetic field (through $\alpha$-$\omega$ dynamo mechanism) is lacking in HAeBe stars and hence the reason for high energy emission is an unsettled problem \citep{1994Appenzeller,2007Mayne,2010Mayne}. 

Many studies have been carried out to understand the origin of X-rays from HAeBe stars. Some of the initial studies to understand the origin of X-ray emission from HAeBe stars were performed by \cite{ZnP1994}, \cite{Damiani1994}, \cite{Skinner_2004} and \cite{Hamaguchi2005} using observations from \textit{ASCA}, \textit{ROSAT}, and \textit{Einstein X-ray Observatory}. These studies could not provide a proper explanation for the origin of X-rays in HAeBe stars due to the limited resolution of the telescope. The advent of the \textit{Chandra X-ray Observatory} (CXO) in 1999 opened a new window (thanks to its high spatial resolution) for studying X-rays from stars. \cite{Stelzer2006} studied a sample of 17 HAeBe stars, of which 13 of them were X-ray emitters. Based on the high plasma temperature obtained from the spectral analysis, this study ruled out radiative winds as the sole emission mechanism and suggested that either a huge fraction of HAeBe stars are intrinsic X-ray emitters or most of the companions to HAeBe stars are yet to be discovered. \cite{Hamidouche2008} studied a sample of 22 HAeBe stars observed with \textit{Chandra} and found X-ray emission from 14 stars, i.e., the sample has a detection rate of $\sim$64\%. They found the X-ray luminosity to be higher than that of TTSs ($10^{30}$ – $10^{31}$ erg s$^{-1}$). Comparing the X-ray properties to other stellar parameters they suggested that the X-ray emission might be intrinsic to HAeBe stars, generated due to star-disk magnetic interaction. Further, \cite{Stelzer2009} studied a sample of 9 HAeBe stars using the CXO with a detection rate of $\sim$100\%. Confirming the results obtained from \cite{Stelzer2006}, \cite{Stelzer2009} suggested that unresolved (multiple stars not resolved in \textit{Chandra}) coronally active low-mass companions as the possible reason for X-ray emission in HAeBe stars. \cite{2021Nunez} studied a small sample of radiative intermediate-mass PMS (R-IMPS) stars and AB-type stars and suggested a similar explanation as \cite{Stelzer2006, Stelzer2009} for the emission of X-rays in HAeBe stars. 

In this work, we present the results obtained by analyzing a sample of HAeBe stars showing X-ray emission and evaluate the possible mechanisms for X-ray emission. Though X-ray emission in TTSs and HAeBe stars has been studied previously, a comparative study with a large sample has not been done, especially with the data from a telescope such as \textit{Chandra} which has an excellent sub-arcsec spatial resolution \citep{2001Tsunemi}. The paper is structured as follows: the details about the data sample of HAeBe stars and TTSs considered for this study are described in Section \ref{section_2:sample}. Section \ref{section_3:dataredalys} describes the data acquisition process and analysis of the X-ray spectra and lightcurves. Section \ref{section_4:results} provides a detailed description of the observed X-ray properties of our sample of HAeBe stars and TTSs in the ONC. Further, the X-ray properties of HAeBe stars and TTSs are compared with their IR and optical properties. In Section \ref{section_5:discuss}, we discuss the role of binaries to understand the X-ray emission from HAeBe stars and explore the possible X-ray emission mechanisms in the light of our study. The main results from this study are concluded in Section \ref{section_6:conclusion}. 

\section{Sample}\label{section_2:sample}
\subsection{\textit{Chandra} data of HAeBe stars}\label{section_2.1:HAeBedat}
We searched for the X-ray observations in the "\textit{Chandra Data Archive}" with a list of 279 HAeBe stars compiled from \citet{Hamaguchi2005}, \citet{Stelzer2006}, \citet{Manoj2006}, \citet{Hamidouche2008}, \citet{Stelzer2009}, \citet{Chen2016}, \citet{Vioque2018}, and identified X-ray observations for 62 HAeBe stars in the field of view (FOV) of Advanced CCD Imaging Spectrometer (ACIS) of \textit{Chandra}. 19 of the 62 sources are presented for the first time in this paper.
They are marked with ``*'' in Table \ref{TableA1:xrayobservations}. 7 of these 19 sources show X-ray emission. 28 of the 62 sources have multi-epoch data. In this paper, for spectral analysis, we consider the observation with the highest exposure and maximum number of counts for stars having multi-epoch data. We also study the temporal behavior of the stars with multi-epoch observations. Of the 62 observations corresponding to 62 sources, 44 are detected in X-rays. Our sample contains a mixture of single and multiple-star systems. The ACIS images of the 44 sources can be classified into 3 categories, i.e., resolved HAeBe star systems (11), unresolved HAeBe star systems (14), and single HAeBe stars (19). We discuss the multiple-star systems to understand the X-ray emission from HAeBe stars and rule out the possibility of TTSs being solely responsible for the observed X-ray emission. Other stellar parameters and the extinction ($A_\mathrm{V}$) along the line of sight are compiled from the literature and tabulated together in Table \ref{TableA1:xrayobservations}. Table \ref{TableA1:xrayobservations} also provides details on the X-ray status of the source, i.e., the sources detected in X-rays are denoted using ``X''. The multiple systems resolved by \textit{Chandra} are denoted using ``X2''. The distances for all the sources are taken from the GAIA EDR3 distance catalog by \cite{BJ2021}.

\subsection{COUP catalog : T Tauri Stars}\label{section_2.2:TTSdat}
The data for TTSs were compiled from the COUP survey program \citep{Getman2005}. The COUP program is the deepest and longest X-ray observation ever made of a young stellar cluster providing a rich and unique data set for a wide range of science studies \cite[e.g.,][]{Getman2005, Preibisch2005, 20051Preibisch, 2005Stelzer}. The CXO was pointed at ONC between 8 January 2003 and 21 January 2003 using the imaging configuration of the ACIS-I, which gives a FOV of 17'x 17'. The total exposure time of the COUP image was 838 ks, corresponding approximately to 9.7 days. The spatial resolution of ACIS is better than 1" over most of the FOV. The very low background allows the reliable detection of sources with as little as $\sim$5 source counts. The final COUP source catalog lists 1616 individual X-ray sources. The ACIS-I chip has a very good point-spread function and a high accuracy of the aspect solution which allows for a clear and unambiguous identification of nearly all X-ray sources with optical or near-infrared counterparts \citep{Preibisch2005}. 
We cross-matched the sources from the COUP catalog \citep{Getman2005}, whose X-ray luminosities are available, with the optical catalog from \cite{DaRio2012} and obtained around 900 sources. The cross-matched list was further filtered to contain only those sources having a membership probability $\geq$ 50\%. These sources are considered to be the \textit{bona fide} members of ONC, thereby reducing the sample size to 716. The number of known populations of stellar cluster members with precisely determined properties significantly increased in this study. So, the number of X-ray emitting TTSs in the sample considered here is more than those listed in \cite{Preibisch2005}.

\section{X-ray data REDUCTION AND ANALYSIS}\label{section_3:dataredalys}
The data reduction was carried out using the CIAO \citep{2006Fruscione} software package version 4.13, in combination with the calibration database (CALDB)\footnote{https://cxc.cfa.harvard.edu/caldb/downloads/index.html} 4.9.4. We followed the standard data analysis procedure recommended by the \textit{Chandra X-ray Center (CXC)}\footnote{\label{2} https://cxc.cfa.harvard.edu/ciao/guides/}. The raw observation files from the archive are reprocessed to obtain the new level-2 event file, which applies the latest calibration and cleans the event file for good time intervals (GTI), removing any spurious or bright pixels (cosmic ray). We then apply an energy filter of 0.3 $-$ 8\,keV to this event file, which is the final event file used for all the scientific analyses. Further, we check for any flaring events in the background lightcurve of the observations. If present then, we follow the standard procedure as recommended by the \textit{CXC}\textsuperscript{\ref{2}} to remove the time bins during which the flaring occurred. This is performed to keep the data uncontaminated due to the background flares. We used the source detection algorithm ``\verb|wavdetect|'' \citep{Freeman2002}, as it is well suited for separating closely spaced point sources. The wavelet scales are set between 1 and 16 in steps of $2^n$ (where, n = {0,1,2,3,4}), and a detection significance of $10^{-6}$ is set to avoid spurious or false detections. This threshold is lowered to $10^{-4}$ for the stars with the closest companions \citep{Stelzer2009}. The position (RA \& Dec) of the sources was then set to the value obtained from the ``\verb|wavdetect|'' algorithm. We then extract the lightcurves and spectra from the event file. For the extraction of the lightcurves and the spectra, we chose a circular region for both the point source and the background. The size of the source region is set to the radius that includes 90\% of the point spread function (PSF) at 1.5\,keV obtained from the \verb|wavdetect| algorithm, and for the background, we choose a source-free region within the same CCD chip. For the sources that are partially resolved such as in Figure \ref{CCD_images}, the extraction regions of the primary and the secondary source were carefully chosen to not overlap with one another using appropriate PSF fraction. In cases where the extraction regions of the primary and the secondary source overlap the common region between them was excluded to minimize the source contamination \citep{2014ApJ...788..101Skinner}. This contamination was not significant in any of the cases.

The targets AK\,Sco, BD+30\,549, BP\,PSc, HD\,100546, R\,CrA, HBC\,442, and HD\,36982 are very faint and have been observed at different epochs. To obtain their spectrum with a better SNR, we co-added the spectra observed at different epochs, using the script ``\verb|combine_spectra|'' (Table \ref{TableA1:xrayobservations}). This was performed only if the sources did not show variability at the different epochs. We checked the variability for these 7 sources by calculating the hardness ratio (HR) at different epochs using Equation \eqref{eq1}. More details with respect to HR are provided in Section \ref{section_3.1:xraylightcurves}.
The sample size in our work is quite large (44 X-ray emitting sources) compared to the previous studies. The X-ray parameters of 37 out of 44 HAeBe stars in our sample have previously been studied in the literature. Further, we found new X-ray detections for 7 HAeBe stars (See Table \ref{TableA1:xrayobservations}). In order to maintain homogeneity with respect to the data analysis, we re-analyzed the X-ray parameters for all the sources in our sample. The X-ray parameters obtained from our analysis agree well with the literature.

\subsection{Temporal Analysis: X-ray Lightcurves} \label{section_3.1:xraylightcurves}
The 0.3 $-$ 8\,keV X-ray lightcurves are extracted for the sources with multiple observations. Only the lightcurves with good time resolution and statistics are used for the scientific analysis. We binned the lightcurves of all the detected sources to a fixed bin time. They are binned carefully to not contain any empty bins. The bin time used for the lightcurve of each source is mentioned in Figures \ref{fig_D1:8_lcs} and \ref{fig_D2:HD_37062}. The HEASARC ftool ``\verb|lcstats|'' is used to measure the ``rms fractional variation'' to check the variability of the lightcurve. The HR for all the X-ray emitting sources in the energy range 0.3 $-$ 1\,keV (S) and 1 $-$ 8\,keV (H) was calculated using equation \eqref{eq1}. 
\begin{equation}
    HR = \frac{H-S}{H+S}
    \label{eq1}
\end{equation}

The following 9 sources, V892\,Tau, HD\,97300, TY\,CrA, HD\,37062, V372\,Ori, V380\,Ori, HR\,5999, HD\,36939 and [DLM2010] EC\,95 show flare-like features \citep{Giardino2004,Hamaguchi2005}. We observe a rapid rise in the count rate of these sources. To confirm this we calculated the HRs and observed a clear hardening. The HR of these stars increases with the increase in the count rate which is not observed for stars without variable/flaring lightcurves. The sources HD\,97300 and EC\,95 are observed only once while the rest of the 7 sources have multiple observations obtained over different epochs. HR\,5999 has 7 multi-epoch observations and it is resolved only in the Obs ID: 8901, where HR\,5999 was the prime target of the observation. HR\,5999\,A is very faint in X-rays, so its lightcurve could not be extracted. Similarly, the star V892 Tau has 2 multi-epoch observations and is resolved only in Obs ID: 3364. The lightcurves showing flaring/variability are presented in Appendix \ref{lightcurves}. More details on the lightcurves are discussed in detail in Section \ref{section_5.1:roleofbin}.

\subsection{Spectral Analysis: The X-ray Spectra} \label{section_3.2:Xrayspectra}
The X-ray spectra of HAeBe stars were extracted using the CIAO tool ``\verb|specextract|''. The Spectrum of each source was binned to a minimum of 5 or more counts/bin. The grouping was chosen based on the quality of the spectrum (i.e., the total number of counts). The sources having total counts < 100 are binned to contain a minimum of 5 counts/bin. For X-ray sources that have the total number of counts $\geq$ 100, we binned the spectrum to contain a minimum of 15 counts/bin. 

The spectral fitting was performed using the tool XSPEC version 12.11.1d \citep{1996ASPC..101...17Arnaud} available with HEASOFT (version 6.30). We used the plasma emission model ``APEC'' \citep{Smith_2001} along with ``TBABS'' for absorption correction. The solar abundances were set to values from \citet{grsa1998} for the plasma model and \citet{Wilms_2000} for the absorption model. We tried fitting the spectrum with single temperature (1T) and two temperature (2T) models, considering four different model combinations on each spectrum i.e., 1T and 2T models with fixed or variable absorption column (N$_\mathrm{H}$) from interstellar dust and the circumstellar material. 
For fits with fixed absorption, the N$_\mathrm{H}$ of the ``TBABS'' component was set to the values derived from the A$_\mathrm{V}$, using the following equation taken from \cite{Bohlin1978},

\begin{equation}
    \frac{N_H}{E(B-V)} = 5.8 \times 10^{21} cm^{-2} mag^{-1}
    \label{eqn2}
\end{equation}

Since our work deals with HAeBe stars, we calculated the N$_\mathrm{H}$ considering an extinction factor R$_\mathrm{V}$=5. A larger R$_\mathrm{V}$ is used as the circumstellar disk of HAeBe stars is dominated by grains larger in size than the average dust grain in the diffuse interstellar medium \citep{Hernandez2004, Manoj2006}.
The column density of the close companions is assumed to be equal to that of the HAeBe stars \citep{Stelzer2006}. Global metal abundances were fixed to 0.3 times the solar value \cite[see,][for details]{2003Imanishi, 2002Feigelson}. For each source, the initial kT and the emission measure (EM) values for model fitting were taken from the literature.

To obtain the values of unabsorbed flux we use the convolution model "\verb|cflux|'' in the energy range 0.3 $-$ 8\,keV. We tried the model fitting with initial values of model parameters from the literature for four different model combinations (fixed and free N$_\mathrm{H}$ for 1T and 2T APEC models) for each source and adopted the fit with the minimum $\chi^2$. For $\chi^2$ statistics the fit quality is determined by null-hypothesis probability (P$_\mathrm{NULL}$). We evaluated the quality of the fit based on the P$_\mathrm{NULL}$ to check how well the spectrum is described by the model and accept only those spectral fits whose P$_\mathrm{NULL}$ > 5\%. First, we try the spectral fitting with the 1T model by fixing N$_\mathrm{H}$ to a value obtained from equation \ref{eqn2}. We consider this model as the best fit if P$_\mathrm{NULL}$ > 5\%, else 1T model with free N$_\mathrm{H}$ is considered. If none of the 1T models gives a best fit, then the 2T model with fixed N$_\mathrm{H}$ is considered. Finally, if none of the models provide the best fit then we use the 2T APEC model with free N$_\mathrm{H}$.
The above method described is used only for $\chi^2$ statistics which works best only for the bright sources (>100 counts) i.e., in our case, the stellar spectrum with at least 15 counts/bin, as this follows a Gaussian distribution \citep{bevington2003data}. The faint sources (<100 counts), i.e., stellar spectrum with less than 10 counts/bin are best explained by a simple Poissonian distribution. Here we used cash statistics (c-statistics) as the test statistic for such data. The quality of the fit for the c-statistics is determined by the ``\verb|goodness|'' command which plots a histogram of probability versus $\chi^2$.
If the statistic of observed data exceeds 90\% of the simulated statistic values, then the model is rejected at 90\% confidence\footnote{https://heasarc.gsfc.nasa.gov/docs/xanadu/xspec/XspecManual.pdf}. The unabsorbed X-ray flux (F$_\mathrm{X}$) is obtained from the model fitting and the unabsorbed X-ray luminosity (L$_\mathrm{X}$) for all the sources is calculated using the formula,

\begin{equation}
    \centering
    L_X = {4 \pi d^2 F_X}
    \label{lumflux}
\end{equation}

where d is the distance to the source. All the X-ray parameters in the model were obtained with a 90 percent confidence range. The X-ray parameters of the sources obtained by the above-mentioned methods agree well with previous studies \citep{Stelzer2006, Stelzer2009, Hamidouche2008}.

The spectrum of sources MWC\,297\,A, MWC\,297\,B, HR\,5999\,A, T\,CrA and R\,Mon having very few counts ($\leq$10 counts) could not be fit with a model. We obtained the count rate of these sources using the CIAO tool ``\verb|aprates|''. The L$_\mathrm{X}$ of these sources was then obtained using ``PIMMS'', assuming an optically thin plasma of 10 MK ($\sim$1\,keV), abundance of $\sim$ 0.3 and the N$_\mathrm{H}$ derived from the A$_\mathrm{V}$. The same procedure was used to estimate the upper limits of the L$_\mathrm{X}$ for the non-detected sources. The L$_X$ of the source R\,CrA (counts > 100) was also obtained using PIMMS as the model parameters could not be constrained. Table \ref{TableB1:xrayparam} gives the X-ray parameters of stars fit with 1T and 2T APEC models for HAeBe sources and their companions. Table \ref{TableB2:xraydark} gives the parameters for the sources that could not be detected in X-rays by \textit{Chandra}. 

\section{RESULTS}\label{section_4:results}
Among our sample of 62 HAeBe stars, 44 of them show X-ray emission. \textit{Chandra} has resolved 11 HAeBe stars among the 44, where the resolved companion is at a position >1". Of the 11 resolved systems, there are 5 stars with sub-arcsec companions. Apart from the resolved systems, there are 14 HAeBe stars with unresolved sub-arcsec companions i.e., companion stars within 1" of the primary which \textit{Chandra} could not resolve. So, in total there are 19 out of 44 unresolved sub-arcsec multiple-star systems where i.) the primary and the secondary are HAeBe stars (e.g., AK\,Sco, V892\,Tau), ii.) the primary and secondary are HAeBe stars with a TTS companion (e.g., R\,CrA, V380\,Ori), and iii.) the primary is a HAeBe star and its companions are TTSs (e.g., TY\,Cra, HD\,97300). In this paper, the group of 19 stars is identified as non-single stars. We evaluated a sub-arcsec companion (stars with separation < 1") fraction of 43.1$\%$ for the stars detected in X-rays in this paper. 25 out of 44 stars (56.8$\%$) are either confirmed to have no stellar companions or have any evidence of the presence of a companion as of yet (we name this group as `single stars'). The details of the companion status of the HAeBe stars in our sample are given in Table \ref{TableC1:individualstars}.

The X-ray parameters (kT, L$_\mathrm{X}$ and N$_\mathrm{H}$) of HAeBe stars in our sample obtained from X-ray spectral analysis are tabulated in Table \ref{TableB1:xrayparam}. To understand the origin of the X-ray emission in HAeBe stars, we compare their X-ray parameters with other stellar parameters such as bolometric luminosity, temperature, infrared excess, and accretion. Further, we compare the properties of HAeBe stars with those of TTSs. The parameters of TTSs in the ONC are obtained from the COUP analysis \citep{Getman2005}, as discussed in Section \ref{section_2.2:TTSdat}. 

\begin{figure}
	\includegraphics[width=\columnwidth]{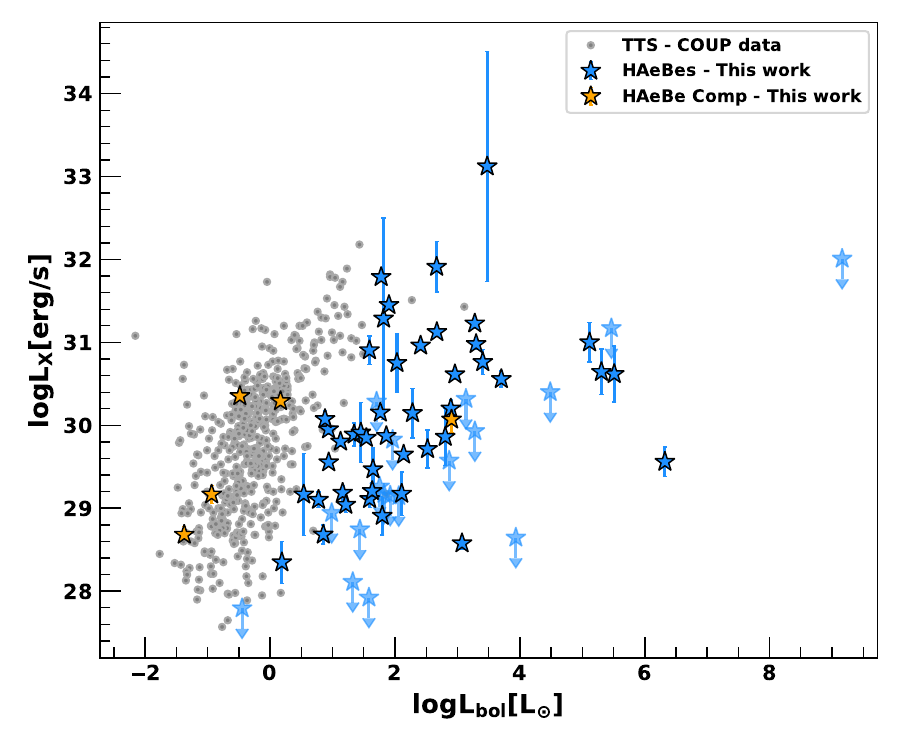}
    \caption{X-ray luminosity (L$_\mathrm{X}$) against bolometric luminosity (L$_\mathrm{bol}$). Grey dots $-$ TTSs (ONC), blue stars $-$ HAeBe stars(our sample), orange stars $-$ close companions to HAeBe stars (our sample), faint blue stars $-$ upper limits (X-ray dark sources).}
    \label{fig_1:LxLb}
\end{figure}
\begin{figure}
	\includegraphics[width=\columnwidth]{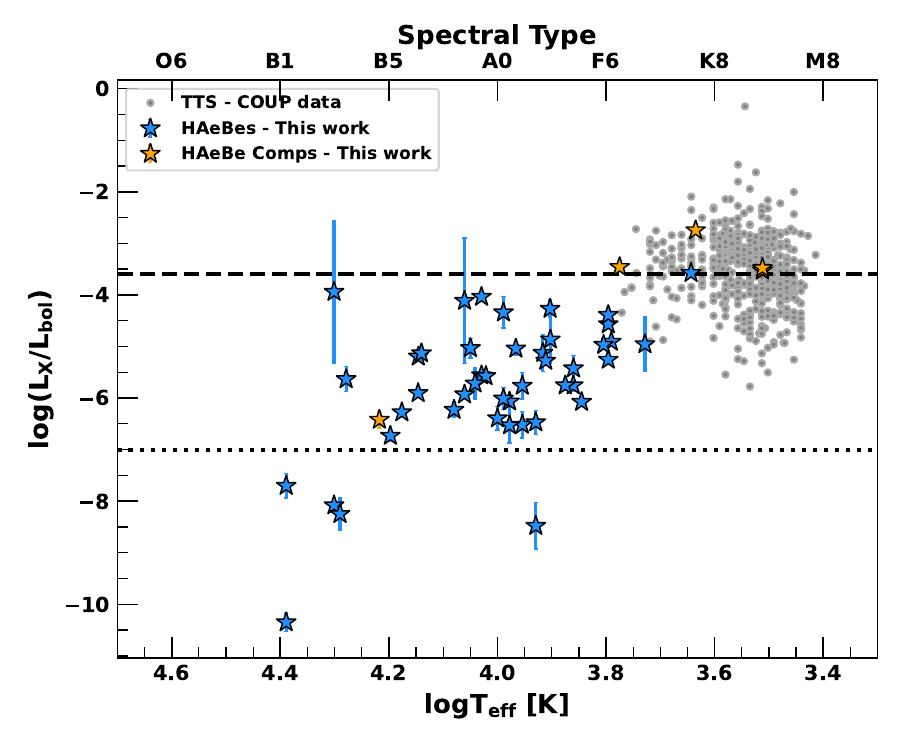}
    \caption{Variation of fractional X-ray luminosity against the temperature and spectral type. The dashed line denotes the mean L$_\mathrm{X}$/L$_\mathrm{bol}$ of TTSs (10$^{-3.6}$) and the dotted line that of main-sequence OB type stars (10$^{-7}$).}
    \label{fig_2:LfTeff}
\end{figure}
\begin{figure*}
	\includegraphics[width=\textwidth]{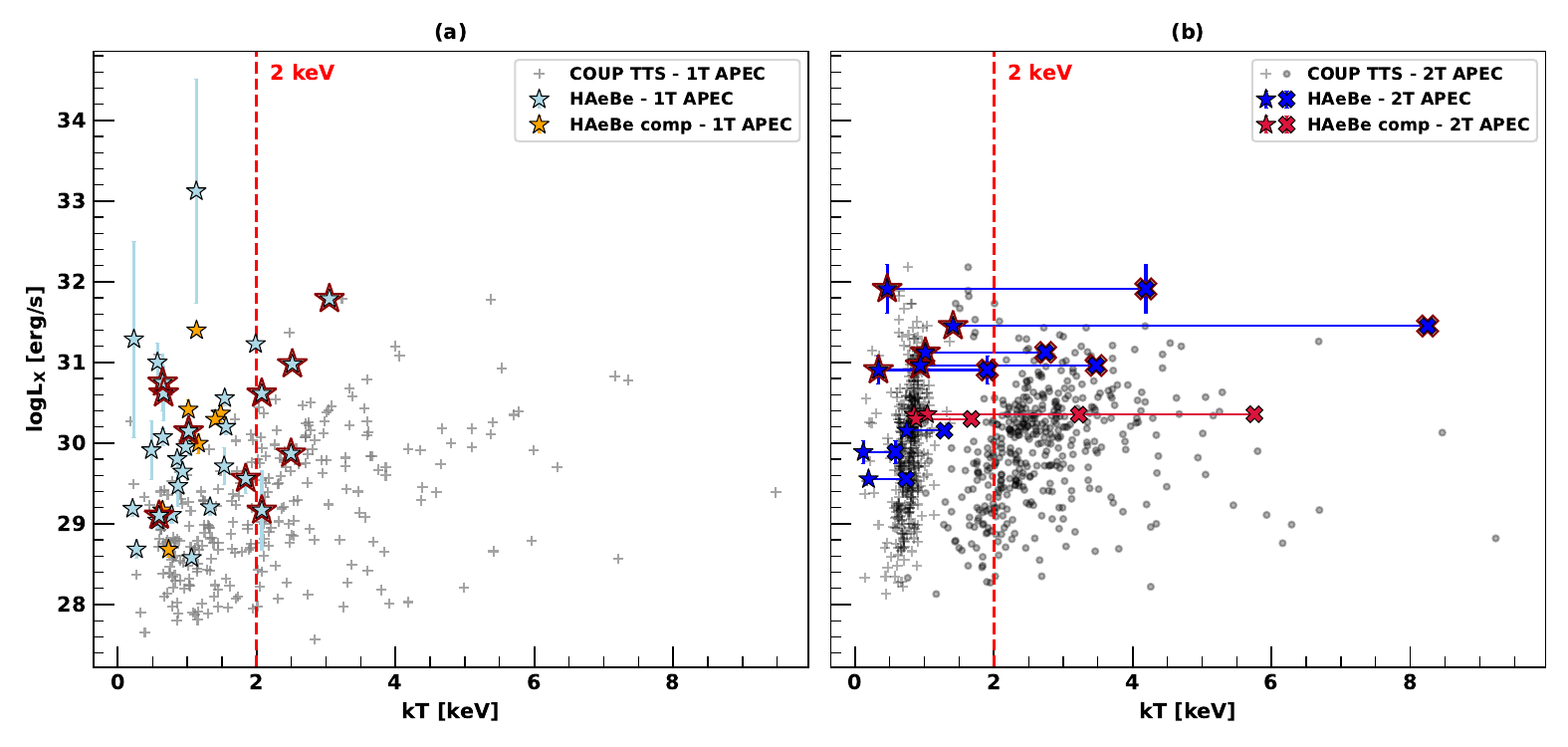}
    \caption{(a). kT vs L$_{X}$ plot for stars fitted with 1T APEC model. The light grey ``+'' symbol denotes TTSs fitted with 1T model and the light blue ``filled star'' symbol denotes HAeBe stars fitted with 1T APEC model. The orange ``filled star'' symbol denotes the resolved companions of HAeBe stars, (b). kT vs L$_{X}$ plot for stars fitted with 2T APEC model. The grey ``+'' and ``o'' symbols denote kT1 and kT2 of TTSs, respectively whereas dark blue ``filled star'' and ``filled cross'' symbols denote kT1 and kT2 of HAeBe stars, respectively. The red ``filled star'' and ``filled cross'' symbols denote kT1 and kT2 of the resolved companions of the HAeBe stars. The blue and red lines connecting these symbols represent the plasma temperatures of their respective stars; The stars outlined in red represent the HAeBe stars with unresolved companions. The red dashed line is marked at 2\,keV denoting the plasma temperature within which single HAeBe stars are found.}
    \label{fig_3:kTvsLx}
\end{figure*}
\subsection{Estimation of L$_\mathrm{X}$/L$_\mathrm{bol}$ of HAeBe stars and their variation with spectral type}\label{section_4.1:LXLbLfTeff}
The variation of X-ray luminosity with bolometric luminosity (L$_\mathrm{bol}$) for HAeBe stars (our sample) and the TTSs from the COUP sample is shown in Figure \ref{fig_1:LxLb}. We compute the bolometric luminosity of HAeBe stars as follows. The observed apparent magnitude (m$_\mathrm{V}$) is corrected for extinction (A$_\mathrm{V}$) and using the distance modulus relation we calculated the absolute magnitude (M$_\mathrm{V}$). The distances to the sources used for the calculation are mentioned in Table \ref{TableA1:xrayobservations}. The m$_\mathrm{V}$ for HAeBe stars were obtained from \cite{Manoj2006}, \cite{Mathew2018}, and \cite{Vioque2018}. From M$_\mathrm{V}$ we estimated the bolometric magnitude (M$_\mathrm{bol}$) using the values of bolometric correction corresponding to each spectral type, from the table listed in \cite{PnM2013}. We then calculate the bolometric luminosity (L$_\mathrm{bol}$) of the HAeBe stars using the equation below,
\begin{equation}
\log\left({\frac{L_{bol}}{L_{\odot}}}\right) = \left({\frac{M_{bol\odot} - M_{bol}}{2.5}}\right)
\label{eqn5}
\end{equation}
where, the bolometric magnitude of the Sun M$_\mathrm{bol,\odot}$ = 4.75. The bolometric luminosity and the unabsorbed X-ray luminosity for the TTSs are obtained from \cite{Getman2005}.

From Figure \ref{fig_1:LxLb}, we see that the L$_\mathrm{bol}$-L$_\mathrm{X}$ distribution of HAeBe stars and TTSs follow a similar trend, with TTSs being more steeper. The X-ray luminosities of both the stellar groups are slightly different, ranging from around $10^{28}$ to $10^{33}$ erg s$^{-1}$ for HAeBe stars and $10^{27}$ to $10^{32}$ erg s$^{-1}$ for TTSs. We see there is a necessity to normalize the L$_\mathrm{X}$, as this gives us an estimate of how bright or faint the HAeBe stars are in X-rays compared to TTSs. Hence, we now consider the plot of L$_\mathrm{X}$/L$_\mathrm{bol}$ against effective temperature/spectral type (Figure \ref{fig_2:LfTeff}). The effective temperature (T$_\mathrm{eff}$) for HAeBe stars were obtained from \cite{Manoj2006}, \cite{Hamidouche2008} and \cite{Vioque2018}. For TTSs, we obtained the T$_\mathrm{eff}$ by cross-matching the final list of TTSs (see, Section \ref{section_2.2:TTSdat}) with \cite{DaRio2012}. The T$_\mathrm{eff}$ was converted to spectral type using the table listed in \cite{PnM2013}.

We see from Figure \ref{fig_1:LxLb} that, although the L$_\mathrm{X}$ of HAeBe stars and TTSs are not substantially different, their L$_\mathrm{X}$/L$_\mathrm{bol}$ appear to greatly differ. It is apparent from Figure \ref{fig_2:LfTeff} that the L$_\mathrm{X}$/L$_\mathrm{bol}$ of HAeBe stars is lower by about 3 to 4 orders of magnitude when compared to that of TTSs. In order to provide a visual comparison in Figure \ref{fig_2:LfTeff}, we have marked the average L$_\mathrm{X}$/L$_\mathrm{bol}$ of TTSs (10$^{-3.6}$) \citep{Preibisch2005} with a horizontal dashed line and that of MS OB stars ($10^{-7}$) \citep{Naze2011} with a horizontal dotted line. The L$_\mathrm{X}$/L$_\mathrm{bol}$ of most of the HAeBe stars in our sample (86\%) fall in between the mean value of TTSs and OB stars. The L$_\mathrm{X}$/L$_\mathrm{bol}$ of HAeBe stars decrease with increasing T$_\mathrm{eff}$, i.e., massive HAeBe stars emit lesser X-ray flux. We do not observe such a trend for TTSs. 

\subsection{Variation of L$_\mathrm{X}$ with kT}\label{section_4.2:XPTkT}
The X-ray spectra of PMS stars could fit well with a 1T or 2T absorbed thermal plasma model \citep{Getman2005}. The plasma temperatures (kT) for the HAeBe stars in our sample were estimated as discussed in Section \ref{section_3.2:Xrayspectra} and it agrees well with the literature \citep{Hamaguchi2005, Stelzer2006, Stelzer2009}. While the coronal temperature is unlikely to be isothermal or be approximated by 1T or 2T models, we still follow this approach as our aim is to homogeneously characterize the plasma temperatures of HAeBe stars in our sample. The plasma temperatures obtained from the model fitting of the X-ray spectrum can be considered as a characteristic temperature \citep{Peres2004, Preibisch2005} and hence we use this parameter to see its variation with L$_\mathrm{X}$. Figure \ref{fig_3:kTvsLx} shows the kT versus L$_\mathrm{X}$ for HAeBe stars (our sample) and TTSs (COUP sample). Figure \ref{fig_3:kTvsLx}(a) represents the HAeBe stars and TTSs that fit with the 1T APEC model and Figure \ref{fig_3:kTvsLx}(b) represents the HAeBe stars and TTSs that fit with the 2T APEC model. In the case of TTSs fitted with 1T model (Figure \ref{fig_3:kTvsLx}a), the plasma temperature of a majority of them is on the hotter side i.e., kT ranging from 0.18 to 9.5\,keV. However, in the case of TTSs with 2T model fits (Figure \ref{fig_3:kTvsLx}b), we observe a cool component (kT1) ranging from 0.14 to 1.37\,keV and a hot component (kT2) ranging from 0.72 to 9.23\,keV \citep{Preibisch2005}. The cool component seems to be constant at $\sim$10 MK (kT $\sim$ 1\,keV), while the hot component seems to increase with increasing L$_\mathrm{X}$. An increase in kT with increasing L$_\mathrm{X}$ is also observed for the TTSs with 1T fit (Figure \ref{fig_3:kTvsLx}a). HAeBe stars with 1T fit which have no known companions have plasma temperatures in the range 0.21 to 1.98\,keV, and HAeBe stars with sub-arcsec companions (stars outlined in red in Figure \ref{fig_3:kTvsLx}a) have plasma temperatures in the range from 0.65 to 3.05\,keV. In the case of HAeBe stars with 2T fit having no known companions, the plasma temperature kT1 ranges from 0.12 to 0.20\,keV and kT2 from 0.58 to 0.74\,keV, whereas for HAeBe stars with sub-arcsec companions (stars outlined in red in Figure \ref{fig_3:kTvsLx}b), the plasma temperature kT1 ranges from 0.34 to 1.41\,keV and kT2 from 1.29 to 8.25\,keV. The sub-arcsec companions to HAeBe stars vary from being low mass to similar mass stars (Table \ref{TableC1:individualstars}). We see hotter plasma temperatures (kT > 2\,keV) for TTSs than that for single HAeBe stars, which are relatively cooler i.e., kT $\leq$ 2\,keV (See Table \ref{TableB1:xrayparam}). The plasma temperatures of HAeBe stars with sub-arcsec companions are hotter (kT > 2\,keV), similar to TTSs. About 50\% of the non-single HAeBe stars fit with the 1T model and 80\% non-single HAeBe stars fit with the 2T model have kT > 2\,keV.

\subsection{Estimation of IR spectral index and its variation with L\textsubscript{X}/L\textsubscript{bol}} \label{section_4.3:IRLf}
It is understood from the infrared studies of HAeBe stars \citep{Manoj2006,Chen2016} that the IR flux excess is due to the presence of dust in their circumstellar disk. \cite{lada1987} showed that the IR spectral index is one of the important criteria used to identify the presence of the disk in YSOs. In this work, we used the IR spectral index $\eta_{(2-4.6)}$ as an indicator of IR excess. A classification scheme for YSOs was quantified by \cite{1994Greene} using the slope of the SEDs in the IR region. The YSOs are distributed into 4 classes, described as Class I ($\eta$ $\geq$ 0.3), Flat (-0.3 $\leq$ $\eta$ < 0.3), Class II (-1.6 $\leq$ $\eta$ < -0.3), Class III ($\eta$ $\leq$ -1.6). 
\begin{figure}
	\includegraphics[width=\columnwidth]{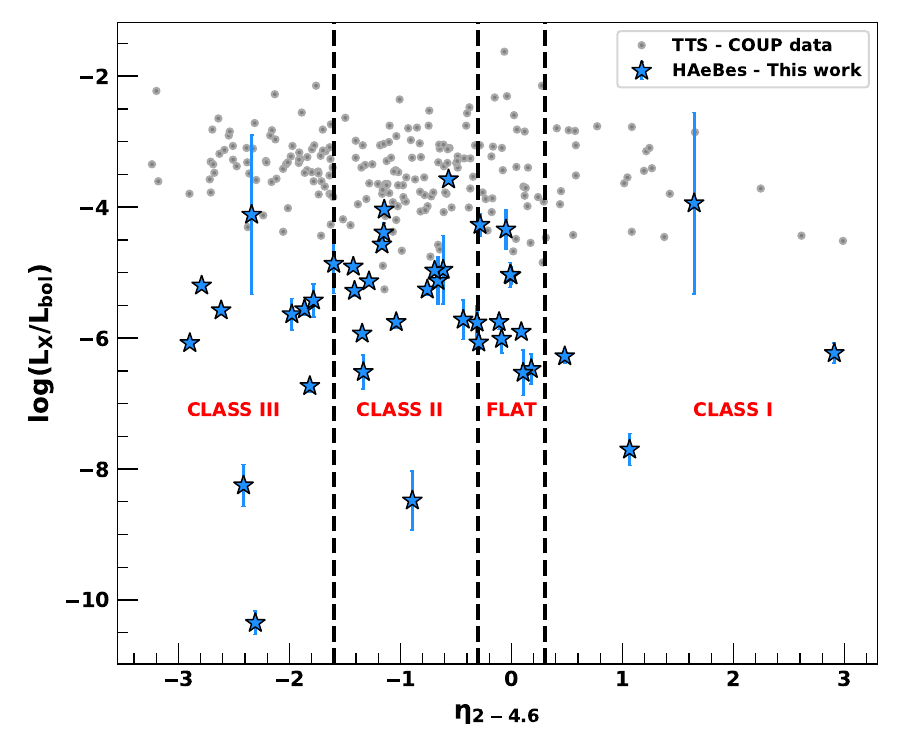}
    \caption{IR Spectral Index against L$_\mathrm{X}$/L$_\mathrm{bol}$ showing the classification scheme by Greene et al. (1994). Class I ($\eta$ $\geq$ 0.3), Flat (-0.3 $\leq$ $\eta$ < 0.3), Class II (-1.6 $\leq$ $\eta$ < -0.3), Class III ($\eta$ $\leq$ -1.6).}
    \label{fig_4:NIRvsLf}
\end{figure}
We obtained the spectral indices in the K$_\mathrm{S}$-W2 IR band for our sample using the method described in \cite{2011Manoj} and \cite{Arun_2019}. The spectral index was calculated using the following equation,
\begin{equation}
    \eta_{\left(\lambda_{1} - \lambda_{2}\right)} = \frac{\log\left({\frac{\lambda_{2}F_{\lambda_{2}}}{\lambda_{1}F_{\lambda_{1}}}}\right)}{\log\left({\frac{\lambda_{2}}{\lambda{1}}}\right)}
\label{eqn6}
\end{equation}
where, $\lambda_{1}$,$\lambda_{2}$ are the two wavelengths considered, and F$_{\lambda_{1}}$, F$_{\lambda_{2}}$ are the flux values at $\lambda_{1}$,$\lambda_{2}$.

Figure \ref{fig_4:NIRvsLf} shows the plot of $\eta_{(2-4.6)}$, an indicator of infrared excess between K$_\mathrm{S}$ and W2 bands, against the L$_\mathrm{X}$/L$_\mathrm{bol}$ for HAeBe stars and TTSs. The excess NIR emission in HAeBe stars is from their inner disk \citep{2004Dullemond, 2004Dullemondb, 2005Dullemond}. Most of the stars in our sample fall under Class II and flat group, which indicates that a majority of the stars in our sample possess a circumstellar disk. We see in Figure \ref{fig_4:NIRvsLf} that similar to TTSs, HAeBe stars do not follow a relation between L$_\mathrm{X}$/L$_\mathrm{bol}$ and $\eta_{(2-4.6)}$. We see that the L$_\mathrm{X}$/L$_\mathrm{bol}$ of both samples does not seem to be affected by the presence or absence of the disk. Further, we need to investigate if the accretion of matter from the circumstellar disk to the star is responsible for the X-ray emission in HAeBe stars, which is evaluated in Section \ref{section_4.4:Halphasec}.

\subsection{Estimation of L$_\mathrm{H\alpha}$/L$_\mathrm{bol}$ and its variation with L$_\mathrm{X}$/L$_\mathrm{bol}$}
\label{section_4.4:Halphasec}
\begin{figure}
	\includegraphics[width=\columnwidth]{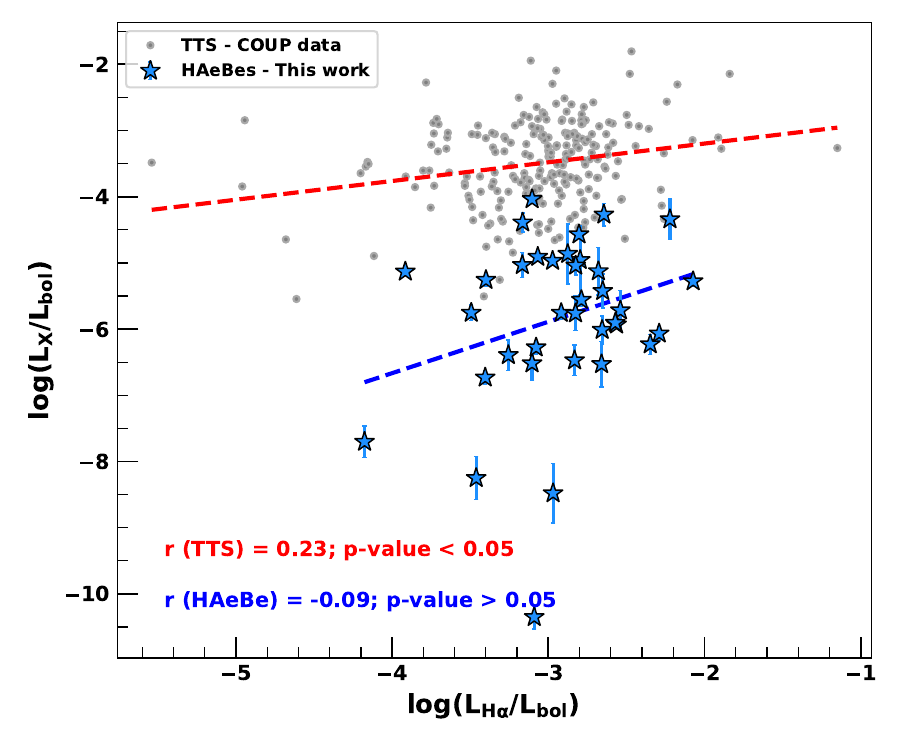}
    \caption{L$_\mathrm{H\alpha}$/L$_\mathrm{bol}$ of HAeBe stars (Blue star) and TTSs (grey dot) plotted against the L$_\mathrm{X}$/L$_\mathrm{bol}$. The Pearson's coefficient for HAeBe stars and TTSs is r = -0.09; p-value = .00051 and r = 0.23; p-value = 0.60167, respectively.}
    \label{fig_5:halpha}
\end{figure}

We know that TTSs accumulate mass through the magnetospheric accretion process. According to this paradigm, the magnetic field lines from the star truncate the disk, and the material from the disk is transferred onto the star. When the in-falling material shocks the photosphere, X-rays are emitted \citep{CnG1998}. It is known that the H$\alpha$ emission line is a good tracer of accretion in PMS stars. Hence, to investigate if the accretion mechanism in HAeBe stars is contributing to the emission of X-rays, we calculate the H$\alpha$ luminosity (L$_\mathrm{H\alpha}$) and plot the L$_\mathrm{H\alpha}$/L$_\mathrm{bol}$ against L$_\mathrm{X}$/L$_\mathrm{bol}$ (Figure \ref{fig_5:halpha}).
We calculated the H$\alpha$ flux using the method described in \cite{Mathew2018}. They used the extinction corrected R-band flux density as a proxy for the continuum flux density underlying the H$\alpha$ line, from which the continuum H$\alpha$ flux (F$_\mathrm{{\nu,cont}(H\alpha)}$) is obtained using the equation \ref{eqn7},
\begin{equation}
    F_\mathrm{{\nu,cont}(H\alpha)} = F_\mathrm{{\nu,0}} \times 10^\mathrm{{\left(\frac{-R_0}{2.5}\right)}}
    \label{eqn7}
\end{equation}
where F$_\mathrm{{\nu,0}}$ = 3.08 $\times$ 10$^{-23}$ W m$^{-2}$ Hz$^{-1}$ and R$_0$ is extinction corrected magnitude. The extinction in R-band (A$_\mathrm{R}$) was estimated from A$_\mathrm{V}$ using the extinction curve of \cite{2009McClure}. From the F$_\mathrm{{\nu,cont}(H\alpha)}$, we estimated the H$\alpha$ line flux using F$_\mathrm{{\lambda,line}(H\alpha)}$ = F$_\mathrm{{\nu,cont}(H\alpha)}$ $\times$ EW$\mathrm{(H\alpha)}$. The absorption corrected equivalent width (EW$\mathrm{(H\alpha)}$) for HAeBe stars was obtained from the literature \citep{Manoj2006, Fairlamb2017, Vioque2018, 2020Wichittanakom}. The H$\alpha$ luminosity (L$_\mathrm{H\alpha}$) is then calculated using the standard flux-luminosity relation, similar to Equation \ref{lumflux}. Figure \ref{fig_5:halpha} shows the plot of L$_\mathrm{H\alpha}$/L$_\mathrm{bol}$ against L$_\mathrm{X}$/L$_\mathrm{bol}$ for both HAeBe stars and TTSs. We perform Pearson's correlation test to check for a linear relation between X-ray emission and the accretion processes. For TTSs, we obtain Pearson's co-efficient r = 0.23 with a p-value = 0.0005 and for HAeBe stars we obtain r = -0.09 with a p-value = 0.6017 at a significance ($\alpha$) level of 0.05. These values suggest that, for TTSs and HAeBe stars, we do not observe any correlation between L$_\mathrm{H\alpha}$/L$_\mathrm{bol}$ and L$_\mathrm{X}$/L$_\mathrm{bol}$. This strongly suggests that the accretion process is not a major contributor to the production of X-rays in TTSs and HAeBe stars.

\begin{figure*}
    \centering
    \subfloat[][]{\includegraphics[width=0.24\textwidth]{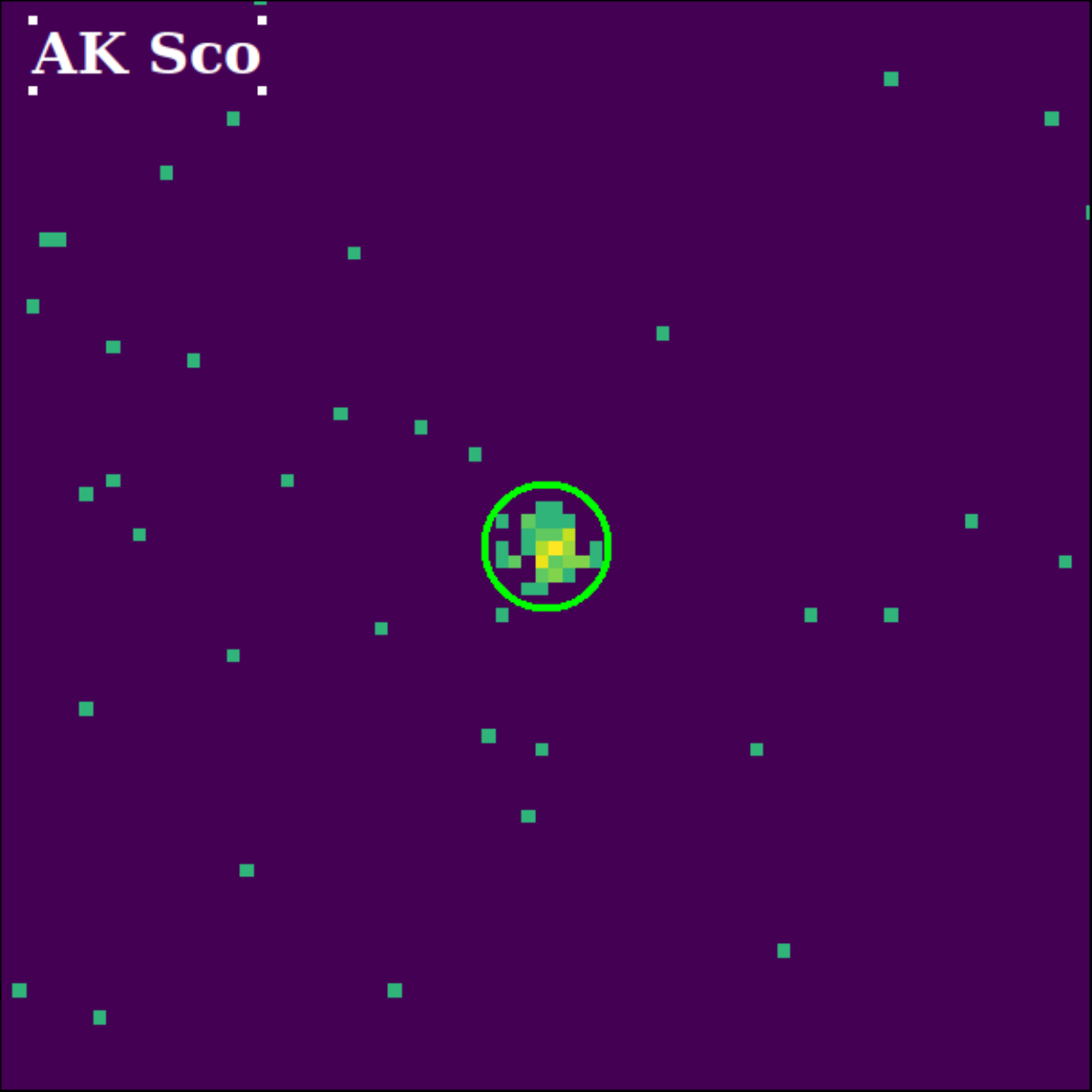}}\hspace{0.1mm} 
    \subfloat[][]{\includegraphics[width=0.24\textwidth]{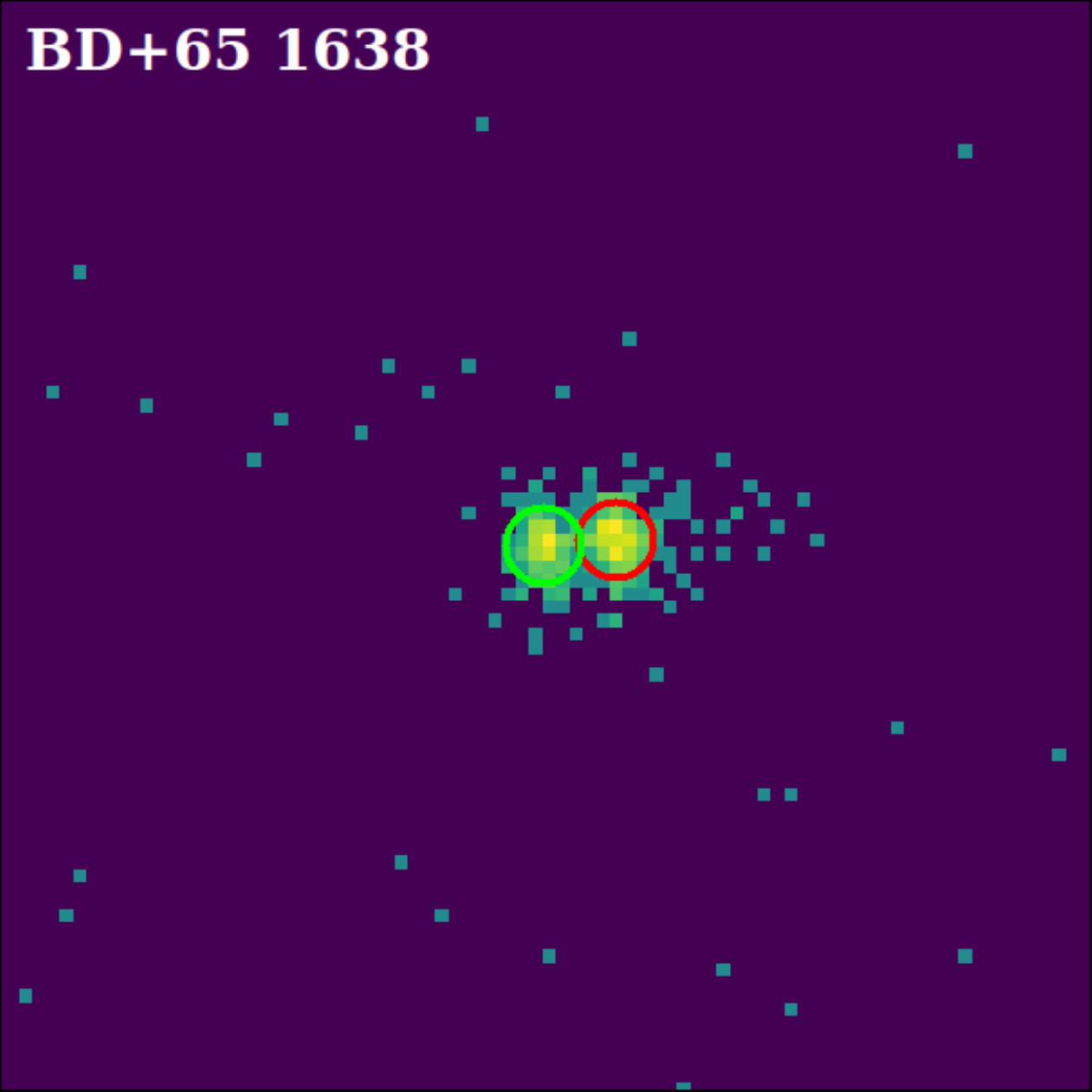}}\hspace{0.1mm} 
    \subfloat[][]{\includegraphics[width=0.24\textwidth]{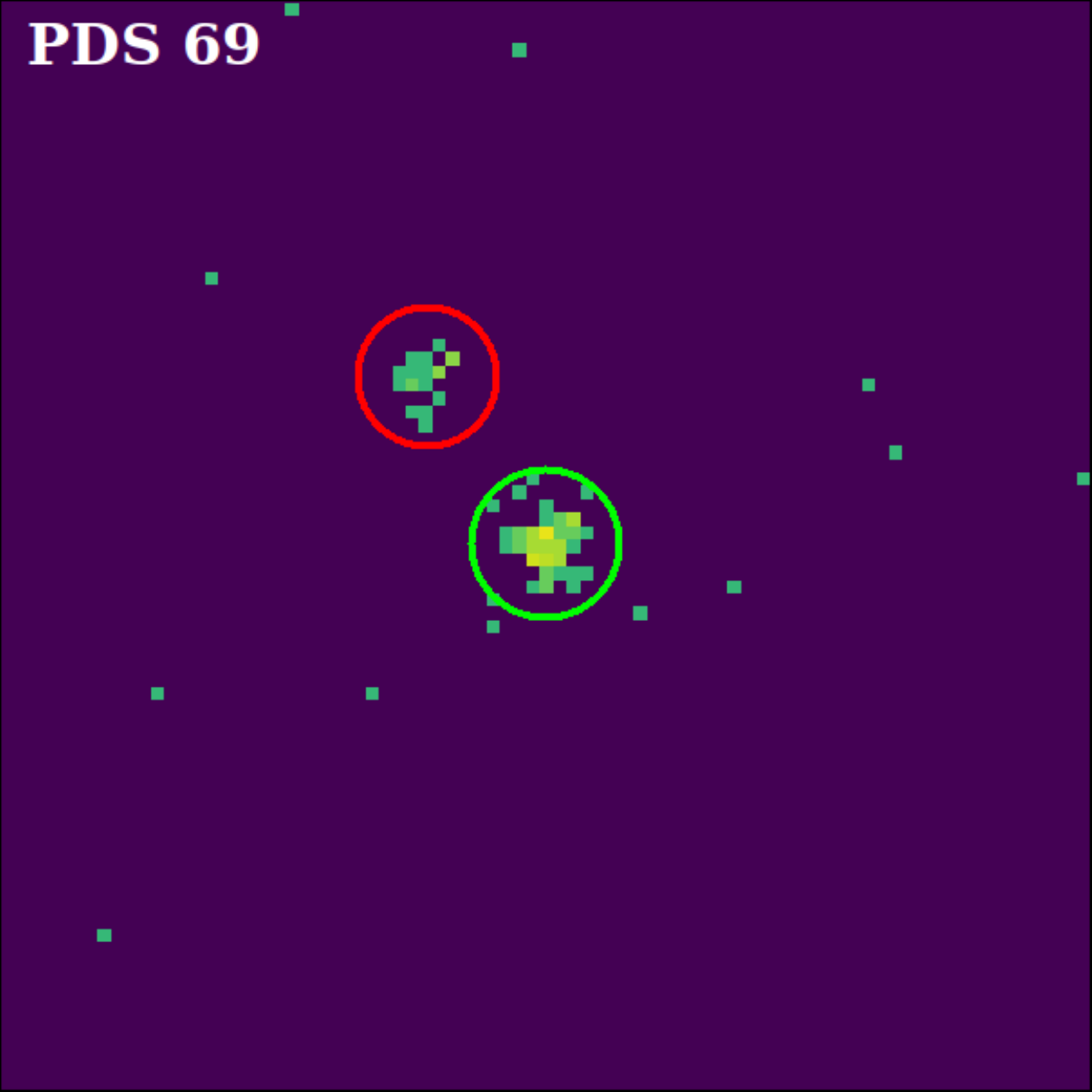}}\hspace{0.1mm}
    \subfloat[][]{\includegraphics[width=0.24\textwidth]{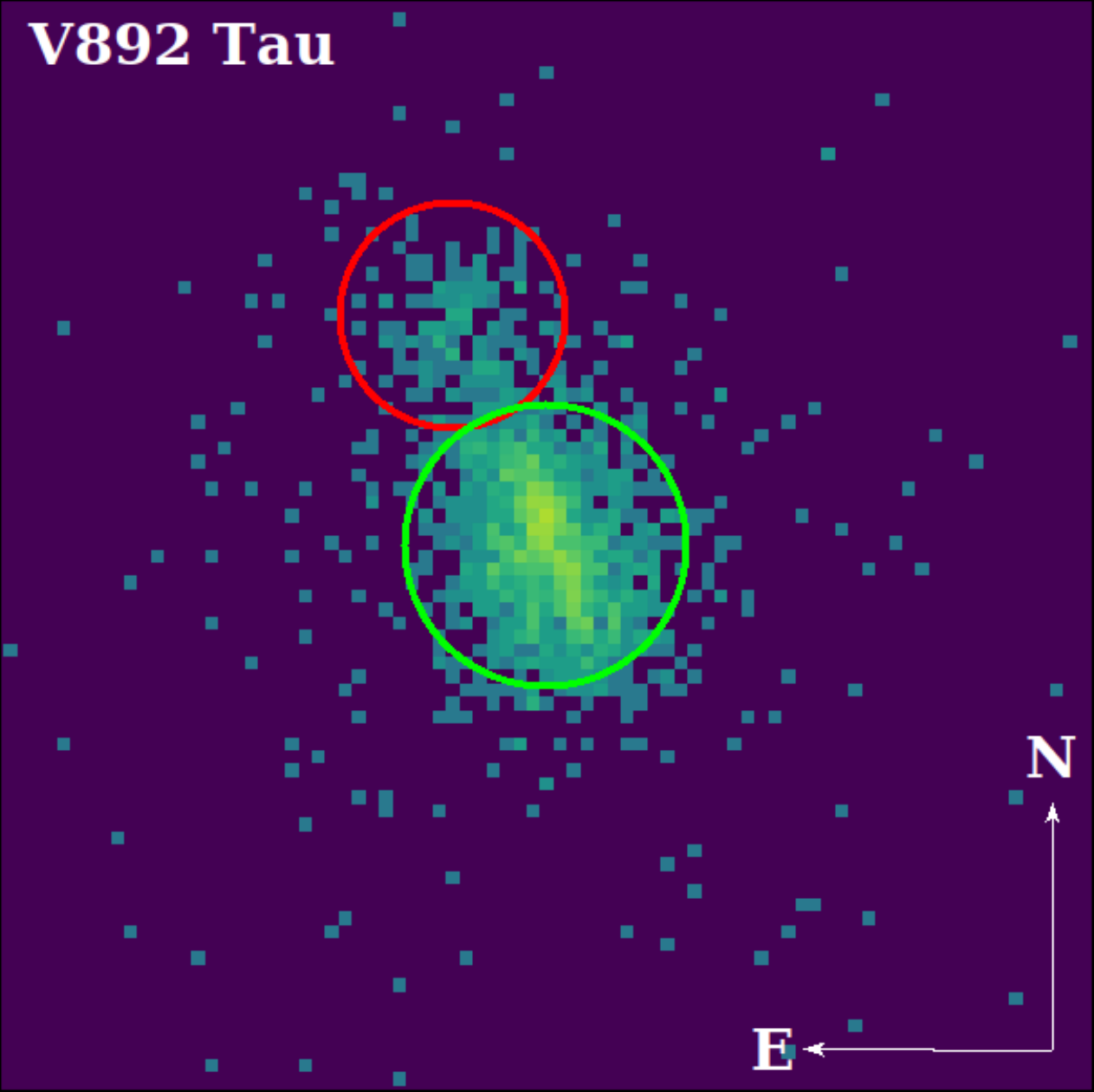}}
    \caption[A]{20" x 20" ACIS $-$ CCD images of the HAeBe stars having equal mass companions (possible HAeBe stars). The images are binned to a pixel size of 0.25". the green circle denotes the photon extraction region of the primary star and the red circle denotes the photon extraction region of the secondary star. For the sources, BD+65\,1638\,A \& B and V892\,Tau\,B, the radius of extraction regions was set to a PSF fraction of 0.84.}
    \label{CCD_images}
\end{figure*}

\section{Discussion}\label{section_5:discuss}
Quite a few mechanisms have been proposed to explain the X-ray emission from HAeBe stars, with the prominent ones being due to the contribution from a low-mass (undetected) companion in the close proximity of the HAeBe star \citep{Stelzer2006, Stelzer2009}, colliding winds producing X-ray emission \citep[similar to what is seen in massive OB-type stars:][]{ZnP1994, Damiani1994}, presence of a shear-powered dynamo in late-type HAeBe stars \citep{TnP1995}, the magnetically confined wind shock model (MCWS) \citep{2005Swartz, Telleschi2007}, and magnetospheric accretion shock model \citep{CnG1998}. From the results obtained in this paper, we propose that the X-ray emission from HAeBe stars has a different origin than TTSs. In the following sub-sections, we rule out the possibility of the presence of an unseen low-mass companion as solely the origin of X-rays and discuss the possible X-ray mechanisms that contribute to the emission of X-rays in HAeBe stars. 

\subsection{Role of binaries}\label{section_5.1:roleofbin} 
In this section, we discuss the details of the various types of companions mentioned in Section \ref{section_4:results} and present three observational details that support intrinsic X-ray emission from HAeBe stars.
\begin{enumerate}
    \item \textbf{Planetary studies -- The presence of planetary companions around the X-ray emitting HAeBe stars:} Some stars in our sample were studied individually using high-resolution instruments and adaptive optics. There were no strong signatures of stellar companions, though planetary companions were discovered. Among the 25 single HAeBe stars, 7 such systems are present. For example, the HAe star AB Aurigae has been studied extensively \citep[e.g.,][]{Boccaletti2020, Currie2022, 2022Zhou}. \cite{Fukagawa2004} studied AB\,Aur in the NIR region using the Coronographic Imager Adaptive Optics systems on the Subaru telescope, in which spiral features within the circumstellar matter were discovered. \cite{2012Tang} suggested the presence of an unseen companion to AB\,Aur at 30 AU as the probable reason for the observed spiral patterns in the disk. A follow-up paper by \cite{2017Tang} identified two spirals in high-resolution ALMA $^{12}$ CO $J$ = 2 $-$ 1 emission observations and suggested that the spiral arms are due to several protoplanets present at 60 $-$ 80 AU (0".4 $-$ 0".6 separation) and another companion at 30 AU (0".16 $-$ 0".18 separation). \cite{Boccaletti2020} presented VLT/SPHERE observations of AB\,Aur and suggested that the two spirals observed in the system are due to the protoplanetary candidates at around 30 AU ($\sim$14$M_\mathrm{Jup}$) and 110 AU ($\sim$3$M_\mathrm{Jup}$) from the center of the star. Similarly, the Herbig Be star HD\,100546 was studied by \cite{2020Perez} using the Atacama Large Millimeter/submillimeter Array. They found evidence for the signatures of 2 protoplanets \citep[$\sim$3.1$M_\mathrm{Jup}$ \& 8.5$M_\mathrm{Jup}$;][]{2021Fedele} and ruled out the possibility of stellar companions by their sparse-aperture masking data. HD 163296 was studied by \cite{2022Kirwan} using spectro-images and position velocity diagrams extracted from the MUSE data cube. They ruled out the presence of any stellar companions within 1 AU $-$ 35 AU and suggested the presence of either a brown dwarf or a massive planet. The details of other stars supporting the evidence of protoplanetary candidates in their inner disk regions are tabulated in Table \ref{TableC1:individualstars} with references. These studies strongly rule out the presence of a stellar companion. Also, the type of X-ray emission observed from the group of stars identified as single stars is towards the cooler side (kT $\leq$ 2\,keV), as discussed in Section \ref{section_4.2:XPTkT}. 
    \item \textbf{Similar mass companions to the X-ray emitting HAeBe stars:} In our sample of stars, there are 4 stars that possess a similar mass/HAeBe star as companions, with no other low-mass companion reported (see, Table \ref{TableC1:individualstars}). Figure \ref{CCD_images} shows the 20" $\times$ 20" ACIS images binned to a pixel size of 0.25" of the HAeBe stars having companions that are possibly HAeBe stars. The details of these 4 stars are discussed below.\\
    AK\,Sco: AK\,Sco (HD\,152404) is a spectroscopic binary \citep[SB2;][]{1989A&A...219..142Andersen} system consisting of two equal-mass (total M$_\star$=2.49M$_\odot$) F5-type stars, with a period of 13.6 days and a large eccentricity (e = 0.47). The distance of separation between the two stellar components is $\sim$ 0.16 AU and they come as close as 11R$_\star$ at the periastron. From the high-resolution studies \citep{2020AJ....160...24Esposito} there has been no detection of another low-mass star in this system. The system has been observed thrice using \textit{CXO}. Due to its very low quality, we combined the data to obtain a good-quality spectrum (see, Section \ref{section_3.2:Xrayspectra}). The details of the X-ray parameters are given in Table \ref{TableB1:xrayparam}. We see a very cool plasma temperature (kT$\sim$0.6\,keV) and no variability in the lightcurve. 
   \\
    BD+65\,1638: BD+65\,1638 is one of the three massive HBe stars in the reflection nebula NGC 7129. It has a spectral type of B2-B3 \citep{1968AJ.....73..233Racine,2003ApJ...592..176Matthews}. \cite{Dham2015} first suggested for the star to be spectroscopic binary. From the source’s L$_\star$ and T$_\mathrm{eff}$, the star is placed very close to the birth-line of a ~6M$_\odot$ object \citep{2003ApJ...592..176Matthews}. In \textit{Chandra} ACIS data the star is resolved into 2 X-ray bright sources with a separation of $\sim$1.3" (Figure \ref{CCD_images}(b)). This system is one of the bright HBe stars in our sample. The X-ray spectra of both the stars fit well with 1T APEC plasma model with plasma temperature kT = 1.98\,keV for BD+65\,1638\,A and kT = 1.13\,keV for BD+65\,1638\,B. The plasma temperature of component A is slightly higher than that of component B. The L$_X$ for both the stars is within the range for B-type stars. We do not observe any variability/flaring in the lightcurve of both the sources. The stellar system seems to be unusually in a quiescent state, as the star has just emerged out of the gaseous envelope.
    \\
    PDS\,69: PDS\,69/H4636 is associated with the reflection nebula NGC\,5367. This is a double-star consisting of two B-type stars of spectral types B7 (PDS\,69\,A) and B4 (PDS\,69\,B). The system is a wide binary at an angular separation of $\sim$3.7". Both the stars show IR excess and the northern component i.e., PDS\,69\,B shows H$\alpha$ in emission \citep{1977MNRAS.180..709Williams,2002Maheswar}. From their positions in the CMD, PDS 69 A and PDS 69 B appear to have masses of about 4.5 and 8 M$_\odot$, respectively \citep{1977MNRAS.180..709Williams}. The wide binary is well resolved in X-rays (\textit{Chandra} data). The ACIS CCD image is shown in figure \ref{CCD_images} (c). The X-ray spectrum is fitted with 1T APEC model for PDS 69 A, with plasma temperature kT = 1.56\,keV and L$_\mathrm{X}$=1.58$\times$10$^{30}$ erg s$^{-1}$. As PDS 69 B is very faint, we used PIMMS to obtain the L$_\mathrm{X}$=1.17$\times$10$^{30}$ erg s$^{-1}$. The X-ray lightcurve of PDS 69 A does not show any signs of variability/flaring or any hard X-ray emission. This double-star system is one example of two resolved HAeBe stars emitting X-rays with no known third unresolved companion.
    \\
    V892\,Tau: V892\,Tau (Elias\,3-1) is a late HAeBe star in the Taurus-Aurigae star-forming region \citep{2005A&A...431..307Smith}. It has a spectral type ranging from B9 to A6 \citep{2009A&A...497..117Alonso-Albi,2016A&A...592A.126vanderMarel}. It is a triple star system, consisting of a central binary of equal mass ($\sim$6M$_\odot$,) separated by $\sim$ 7AU and an M5-type TTS at $\sim$500 AU ($\sim$ 4") from the central binary \citep{2021Long}. We denote the central binary as V892\,Tau\,A and the M-type companion as V892\,Tau\,B. The star has been studied previously in X-rays by \cite{Stelzer2006} and \cite{Giardino2004}. We analyzed the \textit{Chandra} data available for 2 epochs. The system could be resolved from the M-type companion only in one of the observations (Obs ID: 3364) as the coordinates of V892 Tau was near the aim-point of the observation (Figure \ref{CCD_images} (d)). In our analysis, the \textit{Chandra} X-ray spectrum was fit with a 2T APEC model (kT1 = 1\,keV \& kT2 = 2\,keV). A flare is observed in the $\sim$ 18ks X-ray lightcurve, from the region of the central binary, and no variability/flare in the position of the M-type low-mass companion was seen. We also analyzed the observation (Obs ID: 2522) where the source was not resolved, and the X-ray lightcurve showed some variability with a count rate similar to that of the central binary in Obs ID: 3364 (see, Appendix \ref{fig_D1:8_lcs}). The origin of the flare seems to be coming from the two nearly equal mass HAeBe stars \citep{Giardino2004}. Similar characteristics are observed in V372\,Ori, which is a spectroscopic binary with an A-type PMS companion. However, many details are not known about this star and it is a good candidate for future studies.
    \item \textbf{HAeBe stars which present flaring/variability:} Appendix \ref{lightcurves} shows the lightcurves of the 9 HAeBe stars stellar systems unresolved with \textit{Chandra}. Of the 44 stars detected in X-rays, we see X-ray lightcurve variability (short term and long term) and flaring in 9 stars (see, Section \ref{section_3.1:xraylightcurves}; Appendix \ref{lightcurves}). 6 of these have low-mass companions (TY\,CrA, V380\,Ori, HR\,5999, HD\,97300, [DLM2010]\,EC\,95 and HD\,37062), 2 have a similar mass companion (V892\,Tau, V372\,Ori) and 1 (HD 36939) have no companions as of now. X-ray flares in low-mass PMS like TTSs are very common and are characterized by rapid enhancement in the count rate which is followed by slow decay \cite{2007Caramazza}. Low-mass PMS stars show thermal emission with a kT $\sim$ 1 -- 8\,keV and a strong variability with occasional rapid flares consistent with the scenario of enhanced solar-type activity, attributable to magnetic dynamo processes \citep{2003Imanishi, Preibisch2005}. Such characteristics are observed in the HAeBe stars with low-mass companions. TY\,CrA shows a large flare lasting for more than 10ks. While V380\,Ori, HR\,5999, HD\,97300, and HD\,37062 also show flare-like properties, their observation does not contain the details of the whole flare (rise and decay). [DLM2010]\,EC\,95, a proto-Herbig star \citep{2010Dzib} shows short term variability. The plasma temperature of these stars is very high, on the hotter side ranging from kT = 2.75 $-$ 8.25\,keV (Figure \ref{fig_3:kTvsLx}a \& b). V892\,Tau, and V372\,Ori are the only two HAeBe stars known to show flaring activity, with a similar flare shape (Figure \ref{fig_D1:8_lcs}). This is unusual for HAeBe stars as they do not possess convective layers like low-mass stars. The plasma temperature of V892\,Tau is very close to what we see for single HAeBe stars in this paper (kT $\sim$ 2\,keV; Section \ref{section_4.2:XPTkT}). V372\,Ori has a plasma temperature (kT1 = 0.94\,keV; kT2 = 3.75\,keV) and flare characteristics as seen in low-mass stars. It is identified as a spectroscopic binary \citep{1991Abt} consisting of a B9.5+A0 stellar system. The lightcurve of HD\,36939 has a very low count-rate ($\sim$0.002 cts/s) compared to other stars, but we see variability in the count-rate in only one (Obs ID: 4395) of the 14 observations (spanned over a period of 13 years). More high-resolution studies are required to understand the properties of these stellar systems.
\end{enumerate}
The above-mentioned are details of types of sub-arcsec companions to HAeBe stars in our sample (i.e., planetary, similar mass and low-mass companions). The temperature structure of single HAeBe stars and HAeBe stars with companions seems to be different. We see that single HAeBe have cooler plasma temperatures compared to the non-single stars. From the discussion, we can see that in HAeBe stars, X-ray emission is not confined to one mechanism. In Section \ref{section_5.3:xraymechanisms}, we discuss the possible emission mechanisms that can explain the origin of X-ray emission in HAeBe stars.

\subsection{Magnetic fields in Herbig Ae/Be stars}\label{section_5.2:mageticfields}
Magnetic field has been detected in very few HAeBe stars. \cite{2009A&A...502..283Hubrig} studied 21 HAeBe stars using FORS1 mounted on the 8m Kueyen telescope of the VLT. They detected magnetic fields for all of the sources in their sample with a confidence level of $\sim$1$\sigma$. Among the 21 HAeBe stars, 10 were detected with a significance level > 3$\sigma$. The magnetic fields varied from less than 10\,G to $\sim$500\,G. On cross-matching the X-ray emitting stars in our sample with \cite{2009A&A...502..283Hubrig}, we find 11 stars (HD\,100453, HD\,100546, HD\,135344B, HD\,144432, HD\,144668, HD\,150193, AK\,Sco, HD\,163296, HD\,169142, HD\,190073, HD\,97300) to possess magnetic field \citep[see, Table 3][]{2009A&A...502..283Hubrig}. \cite{2009Alecian} studied the multiple-star system V380 Ori using high-resolution spectropolarimeters ESPaDOnS, installed on the 3.6\,m CFHT, and Narval, installed on the 1.9\,m TBL. They measured a magnetic field strength of $\sim$500\,G for the primary. No magnetic field was detected for the secondary component. \cite{2011AN....332.1022Hubrig} studied HD\,31648 using high-resolution polarimetric spectra obtained with SOFIN spectrograph installed at Nordic Optical Telescope. A variable longitudinal magnetic field of the order of a few hundred Gauss up to one kG (-952$\pm$177\,G) was measured. Another study by \cite{2013AN....334.1093Hubrig}, measured magnetic fields on 6 HAeBe stars using the HARPS in spectropolarimeter mode. They measured a weak magnetic field of $\sim$65\,G for HD\,104237. \cite{2015A&A...584A..15Jarvinen}, was able to measure the magnetic field of the primary ($\sim$13\,G) and the secondary ($\sim$129\,G) of the HD\,104237 system using HARPS spectrograph. The primary is a HAe star and the secondary is a TTS (Table \ref{TableC1:individualstars}). \cite{2008MNRAS.387L..23Petit} detected a magnetic field strength of -240$\pm$70\,G was detected for HD\,36982. This was further confirmed by \cite{2013MNRAS.429.1001Alecian} with moderate detection. We see that the Herbig Ae/Be stars possess much weaker magnetic fields than their lower mass counterpart T Tauri stars which have magnetic fields in kG \citep{2019MNRAS.489..886Jarvinen}. The 15 HAeBe stars mentioned in the above discussion emit X-rays whose origin is intimately linked to the presence of the magnetic field. Of these 15 stars, 11 are single HAeBe stars and 4 are non-single HAeBe stars. We find that from our sample of 25 single HAeBe stars, 11 are confirmed to have magnetic fields. Apart from these 11 confirmed magnetic field detections, magnetic fields were measured on 4 single HAeBe stars. However, only upper limits were obtained. For the remaining 10 single HAeBe stars magnetic field detections have not been studied. Among the 19 non-single HAeBe stars, 4 stars are confirmed to have magnetic field detections. Upper limits were measured for 3 non-single HAeBe stars. The magnetic fields of the remaining 12 non-single HAeBe stars have not yet been studied. Further studies on measuring the magnetic fields of X-ray-emitting HAeBe stars would provide more insights. 

\subsection{Possible X-ray emission mechanisms in HAeBe stars}\label{section_5.3:xraymechanisms}
Sun-like stars contain outer convection zones and the X-rays from the corona are linked to a solar-like dynamo. A similar mechanism is thought to be the origin of X-rays in TTSs \citep{Preibisch2005, 2007Telleschi}. Magnetospheric accretion and coronal X-ray are mainly the reason for X-ray emission in TTSs. Massive OB-type stars produce X-rays from radiation-driven instability shocks and magnetically confined wind shocks \citep{1997B&M, Zhekov2007, 2007Montmerle}. Radiative wind models are also ruled as the temperatures of HAeBe stars are not hot enough for the winds to be generated. From the results obtained in this study, we see that the L$_\mathrm{X}$ with respect to the L$_\mathrm{bol}$ decreases towards earlier type stars (Figure \ref{fig_1:LxLb} \& \ref{fig_2:LfTeff}), suggesting a decrease in the X-ray activity levels towards early-type stars. \cite{2020Wichittanakom} showed that the mass-accretion rates and accretion luminosity of HAeBe stars increase with the increase in mass. Our studies show that L$_\mathrm{H\alpha}$, which is a good tracer of mass accretion rate, does not correlate with the L$_\mathrm{X}$ of HAeBe stars, strongly suggesting that in-falling material from the disk is not responsible for X-ray emission (Figure \ref{fig_5:halpha}). \cite{Stelzer2006} and \cite{Stelzer2009} suggest that X-ray emission in HAeBe stars could be attributed to hidden low mass companions, as the binarity rate of their sample was high and also the X-ray properties were very similar to that of other low-mass stars. In Section \ref{section_5.1:roleofbin}, we have documented high-resolution studies that confirm the presence of planetary companions rather than stellar companions in our HAeBe sample (Table \ref{TableC1:individualstars}). Similarly, for HAeBe stars with confirmed unresolved low mass companions, some of which are observed at multi-epochs, we observe long-term and short-term variability with flaring in their X-ray lightcurves. This result along with the soft/cool X-ray temperature of single HAeBe stars and similar mass HAeBe systems suggests that the X-ray emission in HAeBe stars may not be solely due to hidden TTSs but is intrinsic to the star itself. But, what could be the emission mechanism involved in the production of X-rays in HAeBe stars? 

\cite{Telleschi2007} and \cite{2009A&A...494.1041Gunther} studied the high-resolution X-ray spectra of HAeBe stars AB\,Aurigae, and HD\,163296, respectively, which are obtained from the XMM-Newton Reflection Grating Spectrometers (RGS) and EPIC instruments. These stars are confirmed to be single and not known to have any sub-arcsec companions from the high-resolution studies discussed in Section \ref{section_5.1:roleofbin}. The X-ray spectra of both stars are very soft with low photoelectric absorption (the spectrum falls off rapidly above 1\,keV) pointing to a cool source. AB\,Aurigae has temperature components of kT1 = 0.21$^{+0.03}_{-0.03}$\,keV, kT2 = 0.6$^{+0.03}_{-0.04}$\,keV, and L$_\mathrm{X}$ = 4 $\times$ 10$^{29}$ erg s$^{-1}$ in the energy range 0.3 $-$ 10.0\,keV \citep{Telleschi2007}. While, HD\,163296 has temperature components kT1 = 0.21$^{+0.03}_{-0.01}$\,keV, kT2 = 0.51$^{+0.1}_{-0.03}$\,keV, kT3 = 2.7$^{+1.5}_{-0.8}$\,keV, and L$_\mathrm{X}$ = 4.1 $\times$ 10$^{29}$ erg s$^{-1}$ in the energy range 0.2 $-$ 8.0\,keV \citep{2009A&A...494.1041Gunther}. About 35\% of the emission measure was found at 0.2\,keV for AB\,Auriage, whereas it is $\sim$ 60\% for HD\,163296. Further, \cite{2013A&A...552A.142Gunther} studied the \textit{Chandra} data of the source HD\,163296, using the same method as in \cite{2009A&A...494.1041Gunther} and obtained similar model parameters with kT1 = 0.19$^{+0.23}_{-0.02}$\,keV, kT2 = 0.60$^{+0.05}_{-0.04}$\,keV, kT3 = 2.0$^{+1.0}_{-0.5}$\,keV, and L$_\mathrm{X}$ = 3.8 $\times$ 10$^{29}$ erg s$^{-1}$ for Obs ID: 3733 and kT1 = 0.16$^{+0.01}_{-0.01}$\,keV, kT2 = 0.60$^{+0.2}_{-0.02}$\,keV, kT3 = 2.4$^{+0.5}_{-0.3}$\,keV and L$_\mathrm{X}$ = 6.3 $\times$ 10$^{29}$ erg s$^{-1}$ for Obs ID: 12359. These values within the error bars are consistent with our study. \cite{Telleschi2007} and \cite{2009A&A...494.1041Gunther} analyzed the density-sensitive He-like line triplets of O {\sc vii}, observed in the high-resolution spectra of AB\,Aurigae, and HD\,163296. The flux ratio of the forbidden (\textit{f}) to intercombination (\textit{i}) lines (i.e., $\mathcal{R} = f / i$ ratio) at 22.1 \AA, and 21.8 \AA, respectively, point to the low-density regions (such as stellar coronae; \textit{n$_e$} < 10$^{11}$ cm$^{-3}$) as the location for the production of majority the of X-ray emission. The lightcurve of AB\,Aurigae shows a period of $\sim$ 42 hours which is consistent with the modulation period of Mg {\sc ii} and He {\sc i} lines formed in the chromosphere. No such period was found in the lightcurve of HD\,163296, but a short-term variability was observed. Both studies reject the accretion shock model and companion hypothesis. The accretion shock model is rejected based on the high \textit{f\,/\,i} line ratios obtained from the triplets which suggest a low-density environment as the formation region of X-rays, whereas accretion shocks are generated in regions close to the stellar photosphere, in the high-density regions. HD\,163296 drives a jet HH\,409, which could be the reason for the excessive soft X-ray emission, while the $\sim$ 2\,keV component could be from the coronae. The above-mentioned characteristics observed in the high-resolution X-ray spectra of the HAeBe stars AB\,Aurigae, and HD\,163296 clearly point to X-ray emission in the low-density regions such as coronae which supports the presence of magnetic fields in these stars. Based on the evidence discussed above, \cite{Telleschi2007} and \cite{2009A&A...494.1041Gunther} suggested either a wind shock model or magnetically confined winds could be the mechanism responsible for the emission of X-rays.

From the above discussion on individual stars AB\,Aurigae, and HD\,163296 and the magnetic field detections of the HAeBe stars in Section \ref{section_5.2:mageticfields}. It is plausible to say that the X-ray emission in HAeBe stars is due to the presence of magnetic fields. However, the origin of the magnetic field remains unknown. In the study by \cite{2009A&A...502..283Hubrig}, the magnetic fields were found to be stronger in younger HAeBe stars. They also observe that the strength of the magnetic field seems to be increasing with an increase in X-ray luminosity for HAe stars in their sample. The HAe stars are young ($\sim$ 1 $-$ 2 Myr), implying that the X-ray emission decays over time as the star reaches the Main Sequence. This agrees well with the model proposed by \cite{TnP1995}, where a non-solar dynamo operates in rapidly rotating A-type stars based on rotational sheer energy. Other possible mechanisms for X-ray emission from magnetic activity could be due to fossil fields or magnetically confined wind shocks \citep{1997A&A...323..121Babel}. The HAeBe stars identified as single stars in this paper (Table \ref{TableC1:individualstars}), are less studied and do not have enough data to understand their environment and companion status. This can be further explored with multi-wavelength photometry and spectrometry along with high-resolution spectro-polarimetric data to better understand the X-ray emission mechanism responsible for these stars. More evidence is required to support the origin of X-ray emission due to magnetic fields in HAeBe stars.

\section{Conclusions}\label{section_6:conclusion}
We present the X-ray analysis of a large sample of HAeBe stars, investigated with the \textit{Chandra} X-ray telescope having a remarkable $\sim$1" spatial resolution. The main results obtained from this study are given below : 

\begin{enumerate}
\item[1.] Out of 62 HAeBe sources, 44 have been detected in X-rays, i.e., a detection rate of $\sim$70.9\% is observed. We present new X-ray detections of 7 HAeBe stars in this work. The L$_\mathrm{X}$ of HAeBe stars ranges between 10$^{28}$ to 10$^{33}$ erg s$^\mathrm{-1}$ slightly higher than TTSs. The fractional X-ray luminosity of HAeBe stars ranges from $\log$(L$_\mathrm{X}$/L$_\mathrm{bol}$) = -3.5 to -10, most of them falling between $\log$(L$_\mathrm{X}$/L$_\mathrm{bol}$) = -4 to -7, which is in between the value observed for TTSs and OB-type MS stars.
\item[2.] In our sample, 25 out of 44 X-ray emitting HAeBe stars (56.2\%) do not have any reported companions (the single status of some of the sources has been confirmed from previous high-resolution imaging studies). Some of the multi-star systems show variability and flaring, usually observed in TTSs. This evidence rules out the presence of low-mass companions as solely being the X-ray emitters in our sample of HAeBe stars. 
\item[3.] The plasma temperatures of HAeBe stars without any known companions are in the soft energy regime (kT $\leq$ 2\,keV), differing from TTSs in which the hot component (kT > 2\,keV) dominates. About 50\% of non-single HAeBe stars fit with the 1T model and 80\% of non-single HAeBe stars fit with the 2T model show hot X-ray emission, i.e., kT > 2\,keV.
\item[4.] We do not find any correlation between the X-ray luminosity and the disk properties i.e., IR excess ($\eta_{2 - 4.6}$) and fractional accretion luminosity (L$_\mathrm{H\alpha}$/L$_\mathrm{bol}$) for the HAeBe stars in our sample, implying that the X-ray emission is not related to the dust content in the disk or accretion shocks.
\item[5.] From the X-ray spectrum and lightcurve analysis of the HAeBe stars in our sample, we see that the stars with confirmed low-mass companions have a plasma temperature on the hotter side following a similar pattern as TTSs. 9 stars are found to show variability/flaring features in the X-ray lightcurve. Such X-ray variability/flaring is not observed in single HAeBe stars and equal mass HAeBe star systems, except V892\,Tau and V372\,Ori.

From the results obtained, we propose that the X-ray emission from HAeBe is intrinsic and not due to the presence of a close low-mass companion. The X-rays from the HAeBe stars need to be studied further in-depth to constrain the emission mechanism. Studies by \cite{Telleschi2007} and \cite{2009A&A...494.1041Gunther} suggest wind shock models and magnetically confined winds as possible mechanisms for X-ray emission. The validation of these models requires the presence of magnetic fields in HAeBe stars. Based on the previous studies on magnetic fields in literature we found that in single HAeBe stars (i.e., stars with no known sub-arcsec companion), structured magnetic fields could be present. This favors the sheer dynamo theory \citep{TnP1995}, fossil field hypothesis, or magnetically confined wind shocks \citep{1997A&A...323..121Babel}. Further studies with multi-wavelength photometry and spectrometry along with high-resolution spectro-polarimetric data are suggested to support the magnetically driven coronal emission model in these stars.
\end{enumerate}

\section*{Acknowledgements}
We thank the referee for providing constructive comments and suggestions that have helped enhance the quality of the paper. This research has made use of data obtained from the Chandra Data Archive and the software provided by the \textit{Chandra} X-ray Center (CXC) in the application package ``CIAO''. We also made use of the software ``XSPEC'' provided by the High Energy Astrophysics Science Archive Research Center (HEASARC), which is a service of the Astrophysics Science Division at NASA/GSFC. We thank the help desk at \textit{Chandra} and HEASARC for clarifying doubts regarding the X-ray data analysis and software issues. This study has used the \textit{Gaia} EDR3 data to obtain the distances for the sample of stars in this paper. Hence, we express our gratitude to the Gaia collaboration. We thank the SIMBAD database and VizieR online library service for helping with the relevant literature survey. We would like to thank our colleagues Dr. Savithri H Ezhikode, Ms. Nidhi Sabu, and Dr. Arun Roy for providing help with the X-ray and optical analysis part of the paper. We acknowledge the support given by the Center for Research, CHRIST (Deemed to be University), Bangalore, India. VJ and SSK thank the Inter-University Centre for Astronomy and Astrophysics (IUCAA), Pune, India, for the Visiting Associateship. 

\section*{Data Availability}
The data used in this work was accessed from the \textit{Chandra} Data Archive (CDA; \url{https://cda.harvard.edu/chaser/mainEntry.do}).



\bibliographystyle{mnras}
\bibliography{mnras_template} 




\onecolumn
\appendix
\section{Tabulated parameters of \textit{Chandra} Observations of Herbig Ae/Be stars}

\begin{small}
\begin{longtable}[c]{@{}cccccccclcl@{}}
\caption{Observation details of Herbig Ae/Be stars}
\label{TableA1:xrayobservations}\\
\toprule
SI No. & Object & X-ray Iden. & Obs ID & Instrument & RA (J2000) & Dec (J2000) & A$_\mathrm{V}^{(a)}$ & Distance(pc)$^{(b)}$ & No. of Obs. & Ref. \\* \midrule
\endhead
\bottomrule
\endfoot
\endlastfoot
1 & AB Aurigae & X & 3755 & ACIS-S & 04:55:45.86 & 30:33:04.49 & 0.39 & $155.05\pm0.81$ & 1 & 10 \\ 
2 & AK Sco & X & combined\footnote{AK Sco: 983+13274+13275} & ACIS-I & 16:54:44.85 & -36:53:18.55 & 0.35 & $139.12\pm0.57$ & 3 & 4,10,23 \\ 
3 & AS 310 & X & 6399 & ACIS-S & 18:33:21.19 & -04:58:05.78 & 6.06 & $2390.61\pm77.58$ & 1 & 13 \\ 
4 & BD+30 549 & X & combined\footnote{BD+30 549: 642+6436+6437} & ACIS-I & 03:29:19.81 & 31:24:56.81 & 2.30 & $284.78\pm1.93$ & 3 & 4 \\ 
5 & AS 477 & X2 & 6401 & ACIS-S & 21:52:34.10 & 47:13:43.60 & 2.00 & $756.28\pm7.05$ & 1 & 13 \\ 
6 & V361 Cep & X & 6400 & ACIS-S & 21:42:50.22 & 66:06:35.08 & 2.78 & $882.79\pm11.53$ & 1 & 13 \\ 
7 & BD+65 1638 & X2 & 6400 & ACIS-S & 21:42:58.84 & 66:06:10.21 & 3.39 & $1150\pm-$ & 1 & 13 \\ 
8 & BP Psc & X & combined\footnote{BP PSc: 10856+8900} & ACIS-S & 23:22:24.71 & -02:13:41.76 & 1.34 & $211.39\pm23.48$ & 2 & 18\\ 
9 & PDS 69 & X2 & 6423 & ACIS-I & 13:57:43.92 & -39:58:47.27 & 2.58 & $674.34\pm7.63$ & 1 & 11 \\ 
10 & V892 Tau & X2 & 3364 & ACIS-S & 04:18:40.62 & 28:19:15.54 & 11.99 & $133.06\pm1.69$ & 2 & 4,10 \\ 
11 & HD 100453 & X2 & 6429 & ACIS-S & 11:33:05.63 & -54:19:28.95 & 0.10 & $103.68\pm0.22$ & 1 & 15 \\ 
12 & HD 100546 & X & combined\footnote{HD 100546: 2403+3427} & ACIS-I & 11:33:25.39 & -70:11:41.43 & 0.44 & $107.98\pm0.42$ & 2 & 4,10 \\ 
13 & HD 104237 & X2 & 7326 & ACIS-S & 12:00:04.99 & -78:11:34.66 & 0.59 & $106.48\pm0.53$ & 6 & 4,10 \\ 
14 & HD 141569 & $-$ & 981 & ACIS-I & 15:49:57.70 & -03:55:17.00 & 0.50 & $111.33\pm0.37$ & 1 & 4 \\ 
15 & HD 144432 & X2 & 6398 & ACIS-S & 16:06:57.93 & -27:43:10.09 & 0.53 & $154.16\pm0.57$ & 1 & 13 \\ 
16 & HD 147889 & $-$ & 618 & ACIS-I & 16:25:24.32 & -24:27:56.60 & 5.35 & $135.65\pm0.42$ & 1 & 4,10 \\ 
17 & HD 150193 & X2 & 982 & ACIS-I & 16:40:17.95 & -23:53:44.96 & 2.08 & $149.89\pm0.41$ & 1 & 4,10 \\ 
18 & HD 163296 & X & 12359 & ACIS-S & 17:56:21.29 & -21:57:22.38 & 0.24 & $100.50\pm0.34$ & 2 & 4,10,20 \\ 
19 & HD 169142 & X & 6430 & ACIS-S & 18:24:29.79 & -29:46:49.52 & 0.60 & $114.44\pm0.30$ & 1 & 7 \\ 
20 & HD 176386 & $-$ & 3499 & ACIS-I & 19:01:38.90 & -36:53:27.00 & 0.61 & $154.37\pm0.64$ & 8 & 9 \\ 
21 & HD 259431 & X2 & 6397 & ACIS-S & 06:33:05.21 & 10:19:20.16 & 2.05 & $642.26\pm11.33$ & 1 & 13 \\ 
22 & HD 31648 & X & 8939 & ACIS-S & 04:58:46.27 & 29:50:36.64 & 0.55 & $155.08\pm1.16$ & 1 & 16 \\ 
23 & HD 97300 & X & 1867 & ACIS-I & 11:09:50.08 & -76:36:47.65 & 2.15 & $188.78\pm1.25$ & 1 & 4 \\ 
24 & V373 Cep & $-$ & 6400 & ACIS-S & 21:43:06.68 & 66:06:54.60 & 5.14 & $870.73\pm14.91$ & 1 & 13 \\ 
25 & LkHa 25 & $-$ & 2550 & ACIS-I & 06:40:44.60 & 09:48:02.00 & 1.98 & $694.02\pm49.67$ & 3 & 4 \\ 
26 & HD 36982 & X & combined\footnote{HD 36982: 4395+3744+4373+4374+4396+3498+17735} & ACIS-I & 05:35:09.84 & -05:27:52.95 & 1.32 & $404.00\pm3.82$ & 54 & 3 \\ 
27 & MR Ori & $-$ & 4373 & ACIS-S & 05:35:16.97 & -05:21:45.33 & 2.58 & $395.34\pm2.49$ & 86 & 10 \\ 
28 & MWC 297 & X2 & 1883 & ACIS-I & 18:27:39.54 & -03:49:51.93 & 12.46 & $407.41\pm5.19$ & 1 & 4,10 \\ 
29 & PDS 144S & $-$ & 11012 & ACIS-S & 15:49:15.32 & -26:00:54.70 & 0.92 & $1174.13\pm119.29$ & 1 & * \\ 
30 & PDS 581 & $-$ & 4504 & ACIS-S & 19:36:18.90 & 29:32:50 & 4.24 & $1574.38\pm208.99$ & 1 & * \\ 
31 & R CrA & X & combined \footnote{R CrA: 19+3499+4475+5402+5403+5404+5405+19709} & ACIS-I & 19:01:53.66 & -36:57:08.01 & 5.45 & $126.02\pm8.19$ & 9 & 6 \\ 
32 & TY CrA & X & 19709 & ACIS-I & 19:01:40.80 & -36:52:34.01 & 2.84 & $159.13\pm4.20$ & 9 & 9 \\ 
33 & T CrA & X & combined\footnote{T CrA: 19+5404} & ACIS-I & 19:01:58.79 & -36:57:50.33 & 3.00 & $130.00\pm0.00$ & 7 & 9 \\ 
34 & HR 6000 & X & 9921 & ACIS-I & 16:08:34.55 & -39:05:34.18 & 0.16 & $157.43\pm1.05$ & 13 & 13 \\ 
35 & HD 37062 & X & 4374 & ACIS-I & 05:35:31.42 & -05:25:16.12 & 0.65 & $405.49\pm6.45$ & 67 & 1,2 \\ 
36 & V372 Ori & X & 6418 & ACIS-I & 05:34:46.96 & -05:34:15.15 & 0.84 & $447.95\pm10.77$ & 7 & 10 \\ 
37 & V380 Ori & X & 12391 & ACIS-S & 05:36:25.40 & -06:42:57.99 & 4.41 & $379.64\pm15.06$ & 2 & 4 \\ 
38 & HR 5999 & X2 & 8901 & ACIS-I & 16:08:34.27 & -39:06:18.34 & 0.54 & $157.67\pm0.72$ & 7 & 13 \\ 
39 & Z CMa & X & 10845 & ACIS-S & 07:03:43.16 & -11:33:06.24 & 7.89 & $639.94\pm222.48$ & 2 & 4,14 \\ 
40 & HBC 217 & X & 2540 & ACIS-I & 06:40:42.18 & 09:33:37.26 & 0.29 & $703.07\pm6.98$ & 7 & 25 \\ 
41 & HBC 442 & X & combined\footnote{HBC 442: 6416+6417+8936} & ACIS-S & 05:34:14.16 & -05:36:54.38 & 0.38 & $383.52\pm2.58$ & 3 & * \\ 
42 & HD 135344 B & X & 9927 & ACIS-S & 15:15:48.44 & -37:09:16.34 & 0.37 & $134.55\pm0.46$ & 1 & 22 \\ 
43 & HBC 222 & X & 2550 & ACIS-S & 06:40:51.20 & 09:44:46.00 & 0.18 & $704.58\pm6.86$ & 3 & 21 \\ 
44 & HD 135344 & $-$ & 9927 & ACIS-S & 15:15:48.90 & -37:08:56.00 & 0.36 & $134.18\pm0.58$ & 1 & * \\ 
45 & LkHa 257 & $-$ & 15723 & ACIS-I & 21:54:18.80 & 47:12:10.00 & 2.28 & $786.10\pm8.13$ & 1 & * \\ 
46 & LkHa 260 & $-$ & 978 & ACIS-I & 18:19:09.40 & -13:50:41.00 & 5.16 & $1556.18\pm89.85$ & 5 & * \\ 
47 & LkHa 339 & $-$ & 12392 & ACIS-I & 06:10:57.80 & -06:14:40.00 & 4.15 & $824.55\pm10.46$ & 1 & * \\ 
48 & MWC 930 & $-$ & 23646 & ACIS-I & 18:26:25.20 & -07:13:18.00 & 14.06 & $7129.11\pm2189.23$ & 1 & * \\ 
49 & RR Tau & $-$ & 8242 & ACIS-I & 05:39:30.50 & 26:22:27.00 & 3.62 & $798.19\pm18.67$ & 1 & * \\ 
50 & GSC 3975-0579 & $-$ & 16309 & ACIS-S & 21:38:08.50 & 57:26:48.00 & 1.26 & $901.56\pm13.67$ & 2 & * \\ 
51 & HD 36939 & X & 4396 & ACIS-S & 05:34:55.29 & -05:30:22.09 & 0.81 & $416.48\pm7.24$ & 45 & 1,2 \\ 
52 & HD 250550 & X & 21185 & ACIS-S & 06:02:00.00 & 16:30:57 & 0.70 & $748.41\pm25.88$ & 1 & * \\ 
53 & R Mon & X & 21183 & ACIS-S & 06:39:09.95 & 08:44:09.55 & 5.66 & $800.00\pm0.00$ & 1 & * \\ 
54 & HD 244604 & X & 21187 & ACIS-S & 05:31:57.23 & 11:17:41.52 & 1.69 & $397.93\pm3.33$ & 1 & * \\ 
55 & V346 Ori & $-$ & 21186 & ACIS-S & 05:24:42.80 & 01:43:48.00 & 0.31 & $336.39\pm2.39$ & 2 & * \\ 
56 &  {[}DLM2010{]} EC 95a & X & 4479 & ACIS-I & 18:29:57.89 & 01:12:46.15 & 36.00 & $429.20\pm2.00$ & 1 & 8 \\ 
57 & T Ori & $-$ & 2567 & ACIS-S & 05:35:50.50 & -05:28:35.00 & 2.54 & $398.95\pm4.90$ & 25 & * \\ 
58 & MWC 953 & X & 2298 & ACIS-I & 18:43:28.42 & -03:46:17.03 & 5.84 & $1985.43\pm68.25$ & 1 & * \\ 
59 & PDS 37 & $-$ & 7519 & ACIS-S & 10:10:00.32 & -57:02:07.35 & 9.37 & $1626.95\pm63.06$ & 1 & * \\ 
60 & HD 190073 & X & 21184 & ACIS-S & 20:03:02.51 & 05:44:16.65 & 0.34 & $824.88\pm21.93$ & 1 & * \\ 
61 & HD 245906 & X & 8242 & ACIS-I & 05:39:30.47 & +26:19:55.15 & 1.11 & $900.58\pm296.13$ & 1 & * \\ 
62 & AFGL 961 & X & 12142 & ACIS-I & 06:34:37.74 & 04:12:44.20 & 32.00 & $1500.00\pm0.00$ & 1 & 21 \\
\bottomrule
\end{longtable}
\end{small}

\footnotesize
``$-$'' in column (3) denotes non-detections.
``*'' in column (11) denotes new stars which are not previously studied. \\
Table \ref{TableA1:xrayobservations}: (a) The A$_\mathrm{V}$ values were compiled from \cite{PreibischSerpenes}, \cite{Hamaguchi2005}, \cite{Stelzer2006}, \cite{Manoj2006}, \cite{Stelzer2009}, \cite{Dham2015}, \cite{Fairlamb2015}, \cite{Mathew2018}, \cite{Vioque2018} and \cite{Arun_2019}.\\
(b) Distance measurements were compiled from \cite{PreibischSerpenes}, \cite{Manoj2006}, \cite{2011Sandell}, \citep{Dham2015} and \cite{BJ2021}.\\
References: (1) \cite{2002Feigelson}, (2) \cite{2003Feigelson}, (3) \cite{Getman2005}, (4) \cite{Stelzer2006}, (5) \cite{2006Flaccomio}, (6) \cite{2006Forbrich}, (7) \cite{Grady_2007}, (8) \cite{2007Giardino}, (9) \cite{2007Forbrich}, (10) \cite{Hamidouche2008}, (11) \cite{2008Getman}, (12) \cite{2008Testa}, (13) \cite{Stelzer2009}, (14) \cite{2009StelzerZCma}, (15) \cite{2009Collins}, (16) \cite{2010ApJ...719.1565Grady}, (17) \cite{2010AJ....140..266Winston}, (18) \cite{2010ApJ...719L..65Kastner}, (19) \cite{2011ApJ...736...25Forbrich}, (20) \cite{2013A&A...552A.142Gunther}, (21) \cite{2013ApJS..209...27Kuhn}, (22) \cite{2014ApJ...780..150McJunkin}, (23) \cite{2016AJ....152..188Getman}, (24) \cite{2017A&A...602A..10Guarcello}, (25) \cite{2019ApJS..244...28Townsley}.
    
\section{X-ray parameters of HAeBe Stars}

\begin{longtable}[c]{@{}lcccccccccc@{}}
\caption{X-ray parameters of HAeBe Stars from the Model Fitting}
\label{TableB1:xrayparam}\\
\toprule

Name & N$_\mathrm{H}$\,[10$^{22}$ cm$^{-2}$] & kT1\,[keV] & kT2\,[keV] & statistic & P$_\mathrm{null}$ & GOF & $\chi^{2}$ & ${\chi_\mathrm{red}}^{2}$ & DOF & $\log$\,L$_\mathrm{X}$\,[erg s$^{-1}$] \\*  \midrule
\endhead
\bottomrule
\endfoot
\endlastfoot
AB\,Aurigae & 0.05 & 0.78$\pm$0.13 & & c & & 53.00\% & & & 10 & 29.11$\pm$0.10  \\
AK\,Sco & 0.04 & 0.59$\pm$0.14 & & c & & 31.00\% & & & 8 & 29.10$\pm$0.09 \\
AS\,310 & 0.70 & 0.57$\pm$0.33 & & c & & 52.00\% & & & 4 & 31.00$\pm$0.24 \\
BD+30\,549 & 2.22$\pm$1.04 & 0.23$\pm$0.14 & & $\chi^2$ & 5.87E-01 & & 1.93 & 0.64 & 3 & 31.29$\pm$1.22 \\
V361\,Cep & 0.32 & 1.54$\pm$0.56 & & c & & 22.00\% & & & 12 & 30.56$\pm$0.10 \\
BP\,Psc & 7.94 & 2.07$\pm$19.15 & & c & & 3.00\% & & & 7 & 29.17$\pm$0.49 \\
HD\,100546 & 0.05 & 1.40$\pm$0.44 & & c & & 9.00\% & & & 9 & 29.21$\pm$0.10 \\
HD\,163296 & 0.18$\pm$0.05 & 0.12$\pm$0.04 & 0.58$\pm$0.04 & $\chi^2$ & 1.40E-01 & & 83.94 & 1.18 & 71 & 29.89$\pm$0.14 \\
HD\,169142 & 0.07 & 0.21$\pm$0.02 & & c & & 39.00\% & & & 13 & 29.19$\pm$0.09 \\
HD\,31648 & 0.52$\pm$0.25 & 0.48$\pm$0.16 & & $\chi^2$ & 1.74E-01 & & 7.69 & 1.538 & 5 & 29.91$\pm$0.36 \\
HD\,97300 & 0.12$\pm$0.06 & 2.50$\pm$0.38 & & $\chi^2$ & 2.45E-02 & & 77.49 & 1.41 & 55 & 29.87$\pm$0.03 \\
HD\,36982 & 0.15 & 1.06$\pm$0.22 & & $\chi^2$ & 2.79E-01 & & 10.96 & 1.22 & 9 & 28.58$\pm$0.07 \\
R\,CrA & 0.63 & & & PIMMS & & & & & & 29.85$\pm$0.06 \\
TY\,CrA & 0.33 & 1.41$\pm$0.15 & 8.25$\pm$3.23 & $\chi^2$ & 2.21E-01 & & 341.2 & 1.06 & 322 & 31.45$\pm$0.02 \\
HR\,6000 & 0.02 & 0.74$\pm$0.09 & 1.29$\pm$0.24 & $\chi^2$ & 2.01E-01 & & 72.15 & 1.15 & 63 & 30.15$\pm$0.02 \\
HD\,37062 & 0.23$\pm$0.03 & 1.01$\pm$0.07 & 2.75$\pm$0.22 & $\chi^2$ & 2.61E-02 & & 177.63 & 1.24 & 143 & 31.12$\pm$0.02 \\
V372\,Ori & 0.10 & 0.94$\pm$0.21 & 3.47$\pm$0.80 & $\chi^2$ & 4.64E-01 & & 71.41 & 1.01 & 71 & 30.96$\pm$0.03 \\
V380\,Ori & 1.05$\pm$0.13 & 0.45$\pm$0.11 & 4.32$\pm$2.95 & $\chi^2$ & 7.46E-02 & & 123.27 & 1.21 & 102 & 31.91$\pm$0.18 \\
Z\,CMa & 0.92 & 0.66$\pm$0.20 & & $\chi^2$ & 2.90E-01 & & 6.17 & 1.23 & 5 & 30.62$\pm$0.34 \\
HBC\,217 & 0.03 & 0.98$\pm$0.11 & & $\chi^2$ & 5.14E-01 & & 6.23 & 0.89 & 7 & 29.95$\pm$0.06 \\
HBC\,442 & 0.04 & 0.86$\pm$0.12 & & $\chi^2$ & 1.73E-01 & & 16.4 & 1.37 & 12 & 29.80$\pm$0.06 \\
HD\,135344B & 0.04 & 0.20$\pm$0.14 & 0.74$\pm$0.07 & $\chi^2$ & 7.94E-01 & & 29.79 & 0.81 & 37 & 29.55$\pm$0.03 \\
HD\,36939 & 0.09 & 0.93$\pm$0.11 & & $\chi^2$ & 2.65E-01 & & 13.46 & 1.22 & 11 & 29.65$\pm$0.06 \\
HD\,250550 & 0.08 & 1.02$\pm$0.54 & & c & & 38.00\% & & & 3 & 30.15$\pm$0.30 \\
HD\,244604 & 0.20 & & & PIMMS & & & & & & 29.18$\pm$0.26 \\
HD\,245906 & 0.13 & 0.65$\pm$0.35 & & $\chi^2$ & 5.92E-01 & & 1.91 & 0.64 & 3 & 30.75$\pm$0.35 \\
R\,Mon & 0.66 & & & PIMMS & & & & & & 30.76$\pm$0.15 \\
AFGL\,961 & 22.89$\pm$15.31 & 1.13$\pm$1.49 & & $\chi^2$ & 4.54E-01 & & 2.62 & 0.87 & 3 & 33.12$\pm$1.39 \\
{[}DLM2010{]}\,EC\,95a & 3.59$\pm$0.18 & 3.05$\pm$0.26 & & $\chi^2$ & 7.90E-01 & & 175.99 & 0.92 & 192 & 31.79$\pm$0.03 \\
HBC\,222 & 0.02 & 0.65$\pm$0.32 & & c & & 16.00\% & & & 9 & 30.08$\pm$0.11 \\
MWC\,953 & 0.68 & & & PIMMS & & & & & & 30.64$\pm$0.27 \\
HD\,190073 & 0.04 & & & PIMMS & & & & & & 29.86$\pm$0.34 \\
T\,CrA & 0.35 & & & PIMMS & & & & & & 28.35$\pm$0.25 \\ 
AS\,477\,A & 0.23 & 1.53$\pm$3.26 & & c & & 13.00\% & & & 4 & 29.71$\pm$0.23 \\
BD+65\,1638\,A & 0.39 & 1.98$\pm$0.59 & & $\chi^2$ & 5.06E-01 & & 9.28 & 0.928 & 10 & 31.23$\pm$0.05 \\
PDS\,69\,A & 0.30 & 1.56$\pm$0.59 & & c & & 41.00\% & & & 10 & 30.20$\pm$0.10 \\
V892\,Tau\,A & 1.06$\pm$0.17 & 2.07$\pm$0.30 & & $\chi^2$ & 1.41E-01 & & 80.61 & 1.19 & 68 & 30.61$\pm$0.06 \\
HD\,100453\,A & 0.01 & 0.27$\pm$0.05 & & c & & 3.00\% & & & 6 & 28.68$\pm$0.12 \\
HD\,104237\,A & 0.36$\pm$0.11 & 0.34$\pm$0.05 & 1.91$\pm$0.30 & $\chi^2$ & 6.03E-02 & & 82.43 & 1.29 & 64 & 30.91$\pm$0.17 \\
HD\,144432\,A & 0.06 & 0.57$\pm$0.14 & & c & & 13.00\% & & & 8 & 29.04$\pm$0.10 \\
HD\,150193\,A & 0.24 & 0.87$\pm$0.49 & & c & & 13.00\% & & & 3 & 29.47$\pm$0.26 \\
HD\,259431\,A & 0.24 & 2.51$\pm$0.91 & & $\chi^2$ & 3.17E-01 & & 10.43 & 1.16 & 9 & 30.98$\pm$0.07 \\
MWC\,297\,A & 1.45 & 1.84$\pm$2.39 & & c & & 2.00\% & & & 8 & 29.56$\pm$0.18 \\
HR\,5999\,A & 0.06 & & & PIMMS & & & & & & 28.91$\pm$0.23 \\ \hline
AS\,477\,B & 0.23 & 1.48$\pm$0.72 & & c & & 47.00\% & & & 14 & 30.37$\pm$0.10 \\
BD+65\,1638\,B & 0.39 & 1.13$\pm$0.10 & & $\chi^2$ & 6.40E-02 & & 26.61 & 1.57 & 17 & 31.40$\pm$0.04 \\
PDS\,69\,B & 0.30 & & & PIMMS & & & & & & 30.07$\pm$0.16 \\
V892\,Tau\,B & 1.39 & 1.16$\pm$0.26 & & $\chi^2$ & 3.84E-01 & & 8.53 & 1.07 & 8 & 29.99$\pm$0.12 \\
HD\,100453\,B & 0.01 & 0.73$\pm$0.22 & & c & & 36.00\% & & & 7 & 28.68$\pm$0.10 \\
HD\,104237\,B & 0.07 & 0.64$\pm$0.15 & & c & & 13.00\% & & & 9 & 29.17$\pm$0.11 \\
HD\,144432\,B & 0.03$\pm$0.02 & 0.85$\pm$0.11 & 3.23$\pm$0.65 & $\chi^2$ & 7.53E-02 & & 102.17 & 1.23 & 83 & 30.35$\pm$0.04 \\
HD\,150193\,B & 0.24 & 1.40$\pm$0.26 & & $\chi^2$ & 3.61E-01 & & 7.69 & 1.10 & 7 & 30.29$\pm$0.07 \\
HD\,259431\,B & 0.24 & 1.02$\pm$0.30 & & c & & 27.00\% & & & 7 & 30.42$\pm$0.12 \\
MWC\,297\,B & 1.45 & & & PIMMS & & & & & & 29.69$\pm$0.21 \\
HR\,5999\,B & 0.06 & 1.04$\pm$0.10 & >5.75 & $\chi^2$ & 1.73E-01 & & 40.51 & 1.23 & 33 & 30.36$\pm$0.05 \\
HD\,176386\,B & 0.07 & 0.88$\pm$0.08 & 1.68$\pm$0.34 & $\chi^2$ & 1.50E-01 & & 100.68 & 1.16 & 87 & 30.30$\pm$0.02 \\
V373\,Cep\,B & 0.60 & & & PIMMS & & & & & & 30.29$\pm$0.18 \\
\bottomrule 
\end{longtable}

\begin{longtable}[c]{@{}clcc@{}}
\caption{Upper-limits for L$_\mathrm{X}$ of non-detected sources in X-rays}
\label{TableB2:xraydark}\\
\toprule
SI. No & Name & N$_\mathrm{H}$\,[10$^{22}$ cm$^{-2}$] & $\log$L$_\mathrm{X}$\,[erg s$^{-1}$] \\* \midrule
\endhead
\bottomrule
\endfoot
\endlastfoot
1 & HD\,141569\,A & 0.06 & < 28.75 \\
2 & HD\,147889 & 0.62 & < 28.65 \\
3 & V373\,Cep\,A & 0.60 & < 29.93 \\
4 & LkHa\,25 & 0.23 & < 29.26 \\
5 & MR\,Ori & 0.30 & < 29.15 \\
6 & PDS\,144 S & 0.11 & < 27.79 \\
7 & PDS\,581 & 0.49 & < 30.40 \\
8 & PDS\,37 & 1.09 & < 31.17 \\
9 & HD\,176386\,A & 0.07 & < 27.92 \\
10 & HD\,135344 & 0.04 & < 28.11 \\
11 & LkHa\,257 & 0.27 & < 29.17 \\
12 & LkHa\,260 & 0.60 & < 30.32 \\
13 & LkHa\,339 & 0.48 & < 29.15 \\
14 & MWC\,930 & 1.63 & < 32.01 \\
15 & RR\,Tau & 0.42 & < 29.58 \\
16 & GSC\,3975-0579 & 0.15 & < 30.29 \\
17 & V346\,Ori & 0.04 & < 28.94 \\
18 & T\,Ori & 0.30 & < 29.83 \\* \bottomrule
\end{longtable}
Table \ref{TableB1:xrayparam}: Column(1) provides the name of the object, Column(2) the absorption column density in 10$^{22}$ cm$^{-2}$ units, Column(3) \& (4) the plasma temperature in keV, Column(5) identifies the test statistic used for model fitting, Column(6)(7)(8)(9) \& (10) gives the null-hypothesis value, goodness of fit, $\chi^2$, ${\chi_\mathrm{red}}^{2}$ and degrees of freedom, finally Column(11) gives L$_\mathrm{X}$ in log scale.
Table \ref{TableB2:xraydark}: ``<'' denotes that the value is upper-limit.

\newpage
\section{Details of the companion status of X-ray emitting stars}

\begin{longtable}[c]{@{}lcccll@{}}
\caption{Details of HAeBe stars detected in X-rays}
\label{TableC1:individualstars}\\
\toprule
Object & SpT & Binary & \begin{tabular}[c]{@{}l@{}}Resolved in \\ \textit{Chandra} (Sep")\end{tabular} & \begin{tabular}[c]{@{}c@{}}Details of companion\end{tabular} & Reference \\* \midrule
\endhead
\bottomrule
\endfoot
\endlastfoot
AB\,Aurigae & A0 & No & $-$ & \begin{tabular}[c]{@{}l@{}}Planetary companion (\textless 1")\end{tabular} & 69, 73 \\ \hline
AK\,Sco & F5 & Yes & No & \begin{tabular}[c]{@{}l@{}}Stellar companion \\ (F5-type; \textless{}1")\end{tabular} & 12, 39, 57 \\ \hline
AS\,310 & B1 & No & $-$ & \begin{tabular}[c]{@{}c@{}}$-$\end{tabular} & 75 \\ \hline
BD+30\,549 & B8 & No & $-$ & \begin{tabular}[c]{@{}c@{}}$-$\end{tabular} &  \\ \hline
V361\,Cep & B2 & No & $-$ & \begin{tabular}[c]{@{}c@{}}$-$\end{tabular} &  \\ \hline
BP\,Psc & G9 & Yes & No & \begin{tabular}[c]{@{}l@{}}Stellar companion (\textless 1")\end{tabular} & 31 \\ \hline
HD\,100546 & B9 & No & $-$ & \begin{tabular}[c]{@{}l@{}}Planetary companion (\textless 1")\end{tabular} & 50, 60, 62, 63 \\ \hline
HD\,163296 & A0 & No & $-$ & \begin{tabular}[c]{@{}l@{}}Planetary companion (\textless 1")\end{tabular} & 47, 70, 71 \\ \hline
HD\,169142 & B9 & No & $-$ & \begin{tabular}[c]{@{}l@{}}Planetary companion (\textless 1")\end{tabular} & 43, 44, 48, 51 \\ \hline
HD\,31648 & A5 & No & $-$ & \begin{tabular}[c]{@{}c@{}}$-$\end{tabular} & 75 \\ \hline
HD\,97300 & B9V & Yes & No & \begin{tabular}[c]{@{}l@{}}Stellar companion\\M3.5-type (\textless 1")\end{tabular} & 7, 36, 37 \\ \hline
HD\,36982 & B2 & No & $-$ & \begin{tabular}[c]{@{}c@{}}$-$\end{tabular} &  \\ \hline
R\,CrA & A5 & Yes & No & \begin{tabular}[c]{@{}l@{}}Stellar companions\\equal mass star ($\sim$2.3{M$_\odot$}; \textless 1")\\ +M-type ($\sim$0".156)\end{tabular} & 54, 55, 58, 66 \\ \hline
TY\,CrA & B9 & Yes & No & \begin{tabular}[c]{@{}l@{}}Quadruple star system\\ K2+F-type (\textless 1.2 AU; <1")\\ M4-type ($\sim$41 AU; $\sim$0.3")\end{tabular} & 4, 13, 21, 27 \\ \hline
T\,CrA & F0 & Yes & No & \begin{tabular}[c]{@{}l@{}}Stellar companion\\ ($\sim$1M$_{\odot}$; < 1")\end{tabular} & 74, 76\\ \hline
HR\,6000 & A1.5 & $-$ & $-$ & \begin{tabular}[c]{@{}l@{}}Unknown companion \\ ($\sim$ 52M$_{Jup}$; <1")\end{tabular} & 53 \\ \hline
HD\,37062 & B4 & Yes & No & \begin{tabular}[c]{@{}l@{}}Triple star system\\ 2 stars (LMS+A-type; \textless 1")\end{tabular} & 8, 46 \\ \hline
V372\,Ori & B9.5 & Yes & No & \begin{tabular}[c]{@{}l@{}}Stellar companion\\ Spectroscopic binary; A0.5-type\end{tabular} & 1,10 \\ \hline
V380\,Ori & A1 & Yes & No & \begin{tabular}[c]{@{}l@{}}Quadruple system\\ TTS+IR source (\textless 1")\\ M5 type (\textgreater 1")\end{tabular} & 26, 42 \\ \hline
Z\,CMa & B5+F5 & Yes & No & \begin{tabular}[c]{@{}l@{}}Stellar companion\\ FU Ori (\textless 1")\end{tabular} & 11, 67 \\ \hline
HBC\,217 & G0 & No & $-$ & \begin{tabular}[c]{@{}c@{}}$-$\end{tabular} & 75 \\ \hline
HBC\,442 & F8 & No & $-$ & \begin{tabular}[c]{@{}c@{}}$-$\end{tabular} &  \\ \hline
HD\,135344B & A0 & No & $-$ & \begin{tabular}[c]{@{}l@{}}Planetary companion (\textless 1")\end{tabular} & 41, 45 \\ \hline
HD\,36939 & B8-B9 & No & $-$ & \begin{tabular}[c]{@{}c@{}}$-$\end{tabular} &  \\ \hline
HD\,250550 & B9 & Yes & No & \begin{tabular}[c]{@{}l@{}}Stellar companion\\ LMS ($\sim$0.8M$_\odot$; \textless 1")\end{tabular} & 59, 72 \\ \hline
HD\,244604 & A0 & No & $-$ & \begin{tabular}[c]{@{}c@{}}$-$\end{tabular} & 75 \\ \hline
HD\,245906 & A6 & Yes & No & \begin{tabular}[c]{@{}l@{}}Stellar companion\\ G5-type (\textless 1")\end{tabular} & 33 \\ \hline
R\,Mon & B8 IIIe & Yes & No & \begin{tabular}[c]{@{}l@{}}Stellar companion\\ TTS (\textless 1")\end{tabular} & 5, 20 \\ \hline
AFGL\,961 & B2-B3/B5 & Yes & $-$ & \begin{tabular}[c]{@{}l@{}}Stellar companion \\ early B-type (\textgreater 1")\end{tabular} & 35 \\ \hline
{[}DLM2010{]}\,EC\,95a & K2 & Yes & No & \begin{tabular}[c]{@{}l@{}}Stellar companion\\ TTS (\textless 1")\end{tabular} & 30 \\ \hline
HBC\,222 & F8 & No & $-$ & \begin{tabular}[c]{@{}c@{}}$-$\end{tabular} & 75 \\ \hline
MWC\,953 & B2Ve & No & $-$ & \begin{tabular}[c]{@{}c@{}}$-$\end{tabular} & 75 \\ \hline
V1295\,Aql & A2IVe & No & $-$ & \begin{tabular}[c]{@{}l@{}}Planetary companion (\textless 1")\end{tabular} & 72 \\ \hline
AS\,477 & A0 & Yes & Yes (1.4”) & \begin{tabular}[c]{@{}l@{}}Stellar companion \\ Visual companion (\textgreater 1")\end{tabular} & 9, 28 \\ \hline
BD+65\,1638 & B2 & Yes & Yes (1.332”) & \begin{tabular}[c]{@{}l@{}}Stellar companion \\ B-type (\textgreater 1")\end{tabular} & 28, 40 \\ \hline
PDS\,69 & B7 & Yes & Yes (3.774”) & \begin{tabular}[c]{@{}l@{}}Stellar companion\\ B4-type (possible HAeBe ; \textgreater 1")\end{tabular} & 2, 15, 22, 32 \\ 
V892\,Tau & A0 & Yes & Yes (4.523”) & \begin{tabular}[c]{@{}l@{}}Stellar companion\\ A-type (\textless 1") + M3-type (\textgreater 1")\end{tabular} & 24, 61 \\ \hline
HD\,100453 & A9Ve & Yes & Yes (1.007”) & \begin{tabular}[c]{@{}l@{}}Planetary companion(\textless 1") \\ + M-dwarf ($\sim$1")\end{tabular} & 18, 29, 49 \\ \hline
HD\,104237 & A4 & Yes & Yes (1.482”) & \begin{tabular}[c]{@{}l@{}}Stellar companion\\ K3 (\textless 1") + M3/M4-type (\textgreater 1")\end{tabular} & 16, 25, 38, 52, 68 \\ \hline
HD\,144432 & A8 & Yes & Yes (1.400”) & \begin{tabular}[c]{@{}l@{}}Stellar companion\\K7+M1-type (\textgreater 1")\end{tabular} & 17, 23, 34 \\ \hline
HD\,150193 & A0 & Yes & Yes (1.406”) & \begin{tabular}[c]{@{}l@{}}Stellar companion\\K4-type (\textgreater 1")\end{tabular} & 2, 14, 44 \\ \hline
HD\,259431 & B6 & Yes & Yes (3.112”) & \begin{tabular}[c]{@{}l@{}}Stellar companion \\ $\sim$0.6 M$_\odot$ source (\textless 1")\\ Visual companion (\textgreater 1")\end{tabular} & 19, 33, 72 \\ \hline
MWC\,297 & B1 & Yes & Yes (3.532”) & \begin{tabular}[c]{@{}l@{}}Stellar companion\\ low mass star (\textless 1"; $\sim$0.1-0.5M$_\odot$)\\ Visual companion (\textgreater 1")\end{tabular} & 56, 64, 72 \\ \hline
HR\,5999 & A7 & Yes & Yes (1.574”) & \begin{tabular}[c]{@{}l@{}}Stellar companion\\ TTS (\textgreater 1")\end{tabular} & 3, 6, 65, 72 \\* \bottomrule
\end{longtable}
\footnotesize References: (1) \cite{1991Abt}, (2) \cite{1993Reipurth}, (3) \cite{Stecklum1995}, (4) \cite{1996Corporon}, (5) \cite{1997Ghez}, (6) \cite{1997Close}, (7) \cite{1997Leinert}, (8) \cite{1999Corporon}, (9) \cite{2002Maheswar}, (10) \cite{2002Manoj}, (11) \cite{2002MG}, (12) \cite{2003Alencar}, (13) \cite{2003Chauvin}, (14) \cite{2003Fukagawa}, (15) \citep{2004MNRAS.355.1272Maheswar}, (16) \cite{2004A&A...427..907Bohm}, (17) \cite{2004A&A...416..647Perez}, (18) \cite{2006Chen}, (19) \cite{2006Baines}, (20) \cite{2006Fuente}, (21) \cite{2007Forbrich}, (22) \cite{2007A&A...466..191Haikala}, (23) \cite{2007A&A...464..687Carmona}, (24) \cite{2008Monnier}, (25) \cite{2008Testa}, (26) \cite{2009Alecian}, (27) \cite{2009Boersma}, (28) \cite{Stelzer2009}, (29) \cite{2009Collins}, (30) \cite{2010Dzib}, (31) \cite{2010ApJ...719L..65Kastner}, (32) \cite{2010Haikala}, (33) \cite{2010Wheelwright}, (34) \cite{2011Muller}, (35) \cite{2011Sandell}, (36) \cite{2012Kospal}, (37) \cite{2013Daemgen}, (38) \cite{2013Garcia}, (39) \cite{2013Gomez}, (40) \cite{Dham2015}, (41) \cite{2016Bae}, (42) \cite{2016Rodriguez}, (43) \cite{2017Fedele}, (44) \cite{2017Monnier}, (45) \cite{2018Cazzoletti}, (46) \cite{2018GRAVITY}, (47) \cite{2018Isella}, (48) \cite{2018Ligi}, (49) \cite{2018Wagner}, (50) \cite{2019Brittain}, (51) \cite{2019Gratton}, (52) \cite{2019Jarvinen}, (53) \cite{2019A&A...623A..72Kervella}, (54) \cite{2019A&A...624A...4Mesa}, (55) \cite{2019Sissa}, (56) \cite{2020UG}, (57) \cite{2020Gomez}, (58) \cite{2020Launhardt}, (59) \cite{2020Laws}, (60) \cite{2021Fedele}, (61) \cite{2021Long}, (62) \cite{2020Perez}, (63) \cite{2021Pyerin}, (64) \cite{2021Sallum}, (65) \cite{2021Panic}, (66) \cite{2021Sandell}, (67) \cite{2022Dong}, (68) \cite{2022A&A...663A..53Godoy}, (69) \cite{Currie2022}, (70) \cite{2022Angelo}, (71) \cite{2022Leiendecker}, (72) \cite{2022Rich}, (73) \cite{2022Zhou}, (74) \cite{2023A&A...671A..82Rigliaco}, (75) \cite{2023AJ....165..135Thomas}, (76) \cite{2023ApJ...951....1Whelan}.

\section{HAeBe Stars: Variable lightcurves}\label{lightcurves}
\begin{figure}
    \includegraphics[width=17cm,scale=1]{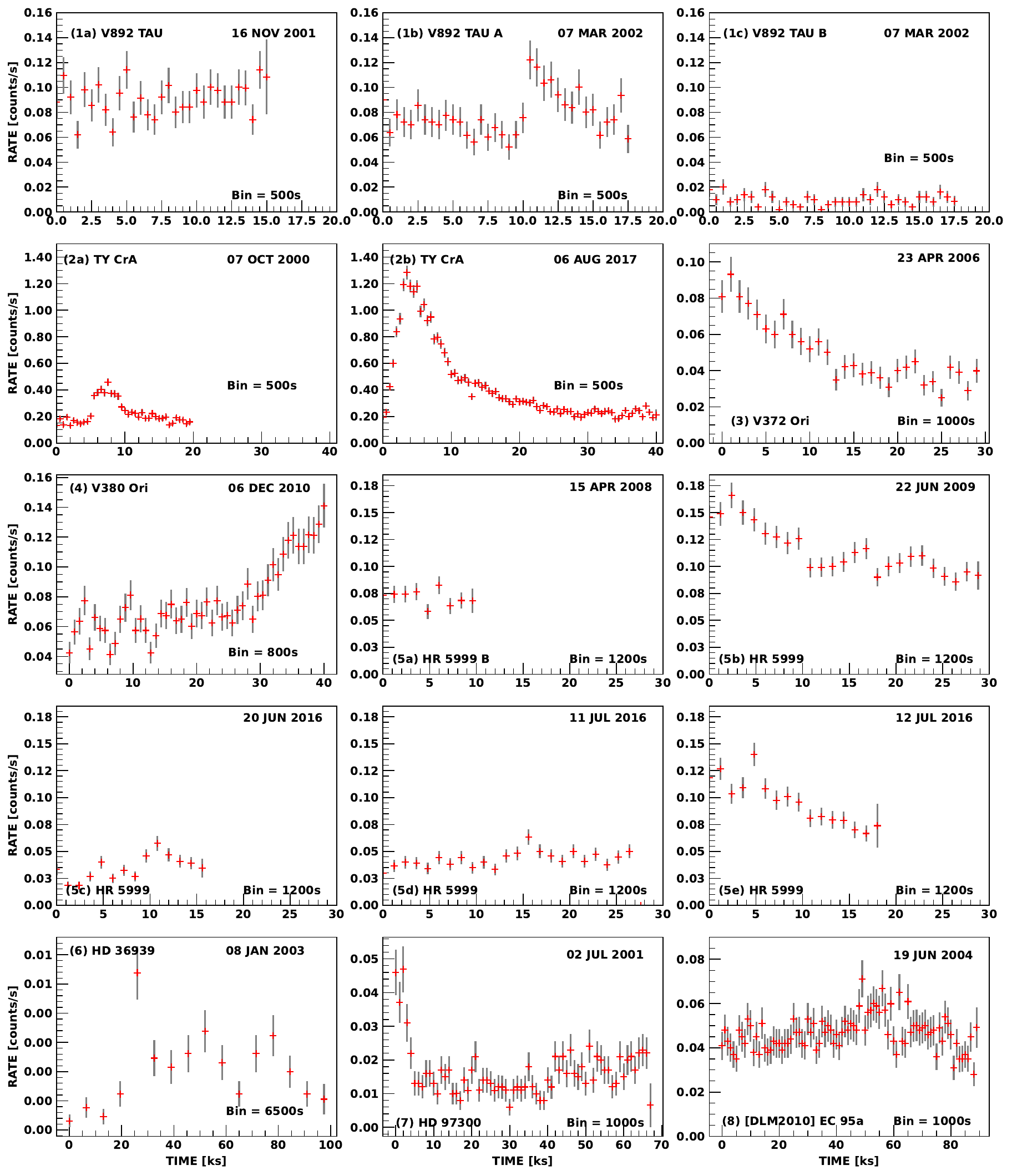}
    \caption{{\textit{Chandra} ACIS lightcurves of 8 stars in the energy range 0.3 $-$ 8.0\,keV that present variability and flaring. The fixed bin time in seconds for each source is mentioned at the bottom right of the plot.}}
    \label{fig_D1:8_lcs}
\end{figure}
\newpage
\begin{figure}
    \includegraphics[width=17cm,scale=1]{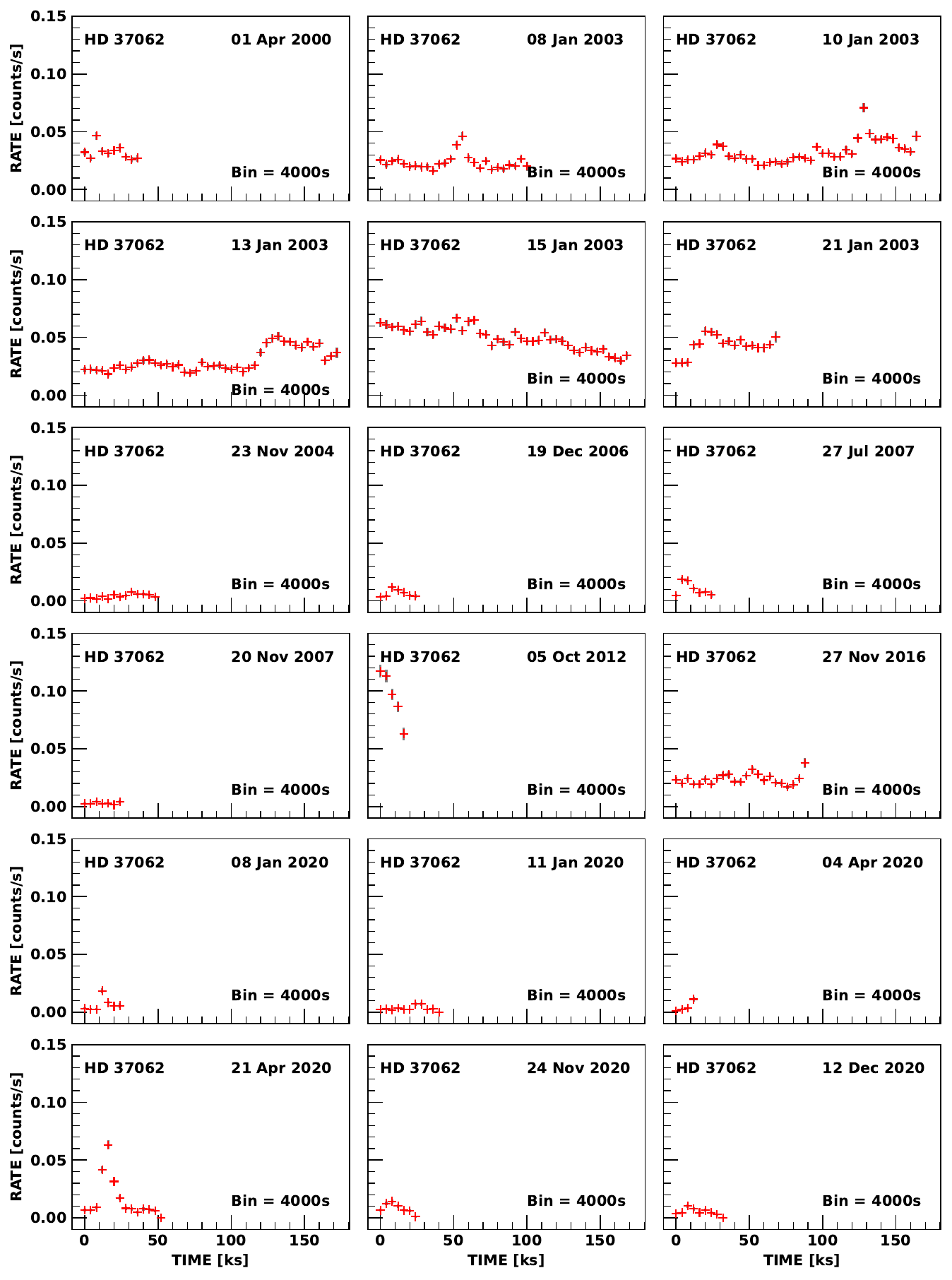}
    \caption{{\textit{Chandra} ACIS lightcurves of HD 37062 in the energy range 0.3 $-$ 8.0\,keV presenting variability and flaring over a period of 20 years. The fixed bin time in seconds for each source is mentioned at the bottom right of the plot.}}
    \label{fig_D2:HD_37062}
\end{figure}



\bsp	
\label{lastpage}
\end{document}